\documentclass[superscriptaddress,twocolumn,english,floatfix,aps,pra,amsmath,amssymb,longbibliography]{revtex4-2}
\usepackage[T1]{fontenc}
\usepackage[utf8]{inputenc}
\usepackage{babel}
\usepackage{comment}
\usepackage{color}
\usepackage{times}
\usepackage{epsfig}
\usepackage{subfigure}
\usepackage{amsmath}
\usepackage{amssymb}
\usepackage{graphicx}
\usepackage[colorlinks=true]{hyperref}
\hypersetup{citecolor=blue,linkcolor=blue,urlcolor=blue}

\begin{document}

\title{Influence of retardation and dispersive surfaces on the regimes of the lateral Casimir-Polder force}

\author{Lucas Queiroz}
\email{lucas.queiroz@ifpa.edu.br}
\affiliation{Instituto Federal de Educa\c{c}\~{a}o, Ci\^{e}ncia e Tecnologia do Par\'{a}, Campus Marabá Rural, 68508-970, Marab\'{a}, Par\'{a}, Brazil}
\affiliation{Faculdade de F\'{i}sica, Universidade Federal do Par\'{a}, 66075-110, Bel\'{e}m, Par\'{a}, Brazil}

\date{\today}

\begin{abstract}
We investigate, by means of the scattering approach, the Casimir-Polder interaction between a neutral anisotropic polarizable particle and a corrugated surface made of a realistic material.
By focusing on the lateral force (arising from the presence of corrugation on the surface), we investigate the conditions for the particle to be attracted to the nearest corrugation peak, valley, or to an intermediate point between a peak and a valley, with such behaviors called peak, valley and intermediate regimes, respectively.
Such regimes of the lateral force were recently predicted in the literature, but in the context of the van der Waals interaction and considering the surface made of some ideal material (a perfectly conducting or a nondispersive dielectric).
Here, we investigate how the occurrence of the mentioned regimes is affected by the consideration of realistic dielectric properties for the surface and also of the retardation in the interaction.
In this context, we show that the consideration of a dispersive surface, when compared to the mentioned idealized materials, can amplify the occurrence of the valley and intermediate regimes.
Moreover, regarding the consideration of retardation, we show that it has a small influence on the occurrence of the valley regime, but, for the intermediate ones, can either amplify or inhibit them.
Such investigation provides a preciser description of the interaction between an anisotropic particle and a corrugated surface, giving a better understanding of the nontrivial aspects of the lateral Casimir-Polder force.
\end{abstract}

\maketitle

\section{Introduction}
\label{sec-intro}

The Casimir-Polder (CP) interaction between a neutral polarizable particle and a surface is a quantum phenomenon that arises due to the quantum fluctuations of the electromagnetic field in vacuum \cite{Casimir-Polder-PhysRev-1948,Milonni-QuantumVacuum-1994}.
In recent years, it has been shown that such interaction could present nontrivial behaviors when considering anisotropic polarizable particles \cite{Levin-PRL-2010, Eberlein-PRA-2011, Buhmann-IJMPA-2016, Abrantes-PRA-2018, Venkataram-PRA-2020, Marchetta-PRA-2021, Bimonte-PRD-2015, Gangaraj-PRB-2018, Antezza-PRB-2020, Nogueira-PRA-2021, Queiroz-PRA-2021, Nogueira-PRA-2022, Queiroz-JPA-2023, Alves-PRA-2023}.
In Ref. \cite{Nogueira-PRA-2021}, specifically, it was investigated the behavior of the lateral van der Waals (vdW) force acting on an anisotropic particle due to the presence of corrugations on the surface, and it was discussed the existence of regimes of this force.
Such regimes are characterized by the possibility of the particle to be attracted towards the nearest corrugation peak, valley, or an intermediate point between a peak and a valley, with such behaviors called peak, valley and intermediate regimes, respectively \cite{Nogueira-PRA-2021}.
In Ref. \cite{Queiroz-PRA-2021}, it was investigated how these regimes are affected by the consideration of dielectric medias, but considering only nondispersive ones, since the focus was on obtaining only a first estimate about the behavior of the regimes in the presence of dielectrics.

The investigations done in Refs. \cite{Nogueira-PRA-2021, Queiroz-PRA-2021} are very useful to obtain preliminary insights about the mentioned regimes.
On the other hand, to obtain more precise results which can be compared with experimental data, it is essential to consider retardation effects and real material properties of the surface.
Thus, in the present paper, we investigate the CP interaction between an anisotropic polarizable particle and a dielectric corrugated surface made of realistic materials.
Following the discussion made in Ref. \cite{Messina-PRA-2009}, we use the scattering approach, at zero-temperature, to calculate the CP energy for this system.
In this way, we generalize a formula found in Ref. \cite{Messina-PRA-2009} in such a way that we include the case of an anisotropic electrically polarizable particle, and study the nontrivial regimes of the lateral force discussed in Refs. \cite{Nogueira-PRA-2021, Queiroz-PRA-2021} in a more general context, showing how they are affected by the consideration of a surface described by a frequency-dependent electric permittivity, as well as retardation effects on the interaction.

The paper is organized as follows. 
In Sec. \ref{sec-used-approach}, we develop the scattering approach to describe the interaction between an anisotropic particle and a corrugated surface, and explore such interaction in the vdW and CP regimes.
In Sec. \ref{sec-sinusoidal}, we apply the obtained formulas to the case of a sinusoidal corrugated surface.
In Sec. \ref{sec-discussions}, we discuss some implications of our results.
In Sec. \ref{sec-final}, we present our final comments.

\section{Interaction energy for an anisotropic polarizable particle and a corrugated surface}
\label{sec-used-approach}

\begin{figure}[h]
\centering
\epsfig{file=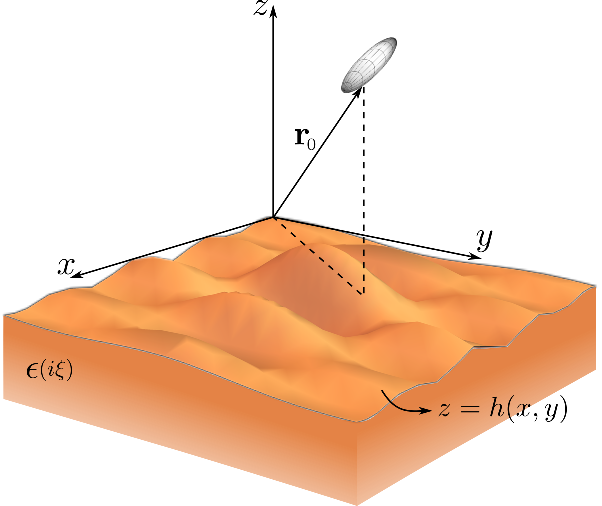, width=0.9 \linewidth}
\caption{
Illustration of an anisotropic polarizable particle (represented by an ellipsoid), located at ${\bf r}_0=x_0\hat{{\bf x}}+y_0\hat{{\bf y}}+z_0\hat{{\bf z}}$, interacting with a general corrugated surface, whose corrugation profile is described by $z=h(x,y)$ [$z_0>h(x,y)$].
}
\label{fig:particula-superficie-geral}
\end{figure}

Let us start considering a neutral anisotropic polarizable particle (here we consider only electrically polarizable ones) in vacuum situated at ${\bf r}_0={\bf r}_{0 ||}+z_0\hat{{\bf z}}$ (with ${\bf r}_{0 ||}=x_0\hat{{\bf x}}+y_0\hat{{\bf y}}$), and interacting with a corrugated surface, as shown in Fig. \ref{fig:particula-superficie-geral}.
The corrugation profile of the surface is described by the function $h(x,y)$ [$z_0>h(x,y)$], which defines a suitable modification of a reference plane at $z = 0$.

By using the scattering approach, in Ref. \cite{Messina-PRA-2009} the authors derived a zero-temperature scattering formula to compute the interaction between an isotropic particle and a corrugated surface.
Here, we take as basis such derivation, but considering an anisotropic particle interacting with the corrugated surface.
This allows us to investigate the nontrivial regimes of the lateral force discussed in Refs. \cite{Nogueira-PRA-2021,Queiroz-PRA-2021} in a more general context, considering realistic dispersive materials to describe the surface, as well as arbitrary particle-surface distances (of course, obeying the restrictions of the discussed case, as shown later).

As in Ref. \cite{Messina-PRA-2009}, it is convenient to derive the scattering formula to the problem discussed here using the plane-wave basis $\left|{\bf k},p\right\rangle $, where ${\bf k}$ represents the lateral wave-vector of a fluctuating electromagnetic field with polarization $p$ (TE for transverse electric and TM for transverse magnetic).
Assuming that $z_0$ is much larger than the dimensions of the particle, the zero-temperature scattering formula to compute the Casimir-Polder interaction is given by \cite{Messina-PRA-2009}
\begin{align}
\nonumber
U\left({\bf r}_{0}\right) & =-\hbar\int_{0}^{\infty}\frac{d\xi}{2\pi}\int\frac{d^{2}{\bf k}}{\left(2\pi\right)^{2}}\int\frac{d^{2}{\bf k}^{\prime}}{\left(2\pi\right)^{2}}e^{-\left(\kappa+\kappa^{\prime}\right)z_{0}}\\
& \times\sum_{p,p^{\prime}}\left\langle {\bf k},p\right|{\cal R}_{S}\left|{\bf k}^{\prime},p^{\prime}\right\rangle \left\langle {\bf k}^{\prime},p^{\prime}\right|{\cal R}_{P}\left|{\bf k},p\right\rangle,
\label{eq:energia-geral}
\end{align}
where $\kappa=\sqrt{\xi^{2}/c^{2}+\left|{\bf k}\right|^{2}}$ is the imaginary $z$-component of the wave-vector associated with the imaginary frequency $\xi$. 
The factor $\exp[-\left(\kappa+\kappa^{\prime}\right)z_{0}]$ represents the round-trip propagation of the field between the surface and the particle.
Finally, ${\cal R}_{P}$ and ${\cal R}_{S}$ are the reflection operators for the particle and the surface, respectively (we remark that these operators are frequency dependent, but to avoid an overload of the notation we are omitting it).

\subsection{The consideration of an anisotropic particle}

The reflection operator for an anisotropic particle can be calculated following the same steps discussed in Ref. \cite{Messina-PRA-2009} in the context of an isotropic one.
The consideration of a ground-state anisotropic particle consists in describing it as an induced electric dipole with a dipole moment given by
\begin{equation}
{\bf d}\left(\omega\right) = \boldsymbol{\alpha}\left(\omega\right) \cdot {\bf E}\left({\bf r}_{0},\omega\right),
\label{eq:dipolo-particula-anisotropica}
\end{equation}
where $\boldsymbol{\alpha}\left(\omega\right)$ is the dynamic polarizability tensor of the particle written in terms of real frequencies $\omega$ (later we consider imaginary frequencies by making $\omega \to i\xi$).
We remark that the polarizability of an anisotropic particle is described by a symmetric tensor of rank two.
For an isotropic particle, this tensor is given by $\boldsymbol{\alpha}=\alpha {\bf I}$ (with ${\bf I}$ being the identity matrix) and thus its polarizability can be simply described by the scalar quantity $\alpha$.
Following the steps discussed in Ref. \cite{Messina-PRA-2009}, but considering Eq. 
\eqref{eq:dipolo-particula-anisotropica}, one obtains 
\begin{align}
\nonumber
\left\langle {\bf k}^{\prime},p^{\prime}\right|{\cal R}_{P}\left|{\bf k},p\right\rangle  & =-\frac{\xi^{2}}{2\epsilon_{0}\kappa^{\prime}c^{2}}e^{i\left({\bf k}-{\bf k}^{\prime}\right)\cdot{\bf r}_{0||}}e^{-\left(\kappa+\kappa^{\prime}\right)z_{0}}\\
 & \times\sum_{m,n}\alpha_{mn}\left(i\xi\right)\left[\hat{\boldsymbol{e}}_{p^{\prime}}^{-}\left({\bf k}^{\prime}\right)\right]_{m}\left[\hat{\boldsymbol{e}}_{p}^{+}\left({\bf k}\right)\right]_{n},
\label{eq:operador-RP}
\end{align}
where $\alpha_{mn}\left(i\xi\right)$ $(m,n=x,y,z)$ are the components of the polarizability tensor $\boldsymbol{\alpha}$, and $\left[\hat{\boldsymbol{e}}_{p^{\prime}}^{-}\left({\bf k}^{\prime}\right)\right]_{m}$ and $\left[\hat{\boldsymbol{e}}_{p}^{+}\left({\bf k}\right)\right]_{n}$ are the components of the unit polarization vectors for the outgoing and the incoming fields, respectively. 
The $\pm$ notation represents the propagation direction of the field along the $z$-axis, and thus the superscript $+$ refers to the incoming field on the particle, and $-$ refers to the outgoing field.
For a field propagating with complete wave-vector ${\bf K}^\pm = {\bf k} \pm k_z \hat{{\bf z}} $, with $k_z = \text{sgn}(\omega) \sqrt{\omega^{2}/c^{2}-\left|{\bf k}\right|^{2}}$ (when we make $\omega \to i\xi$, we get $k_z \to i\kappa$), the unit polarization vectors corresponding to the TE and TM polarizations are defined as
\begin{align}
\hat{\boldsymbol{e}}_{\text{TE}}^{+}\left({\bf k}\right) & =\hat{\boldsymbol{e}}_{\text{TE}}^{-}\left({\bf k}\right)=\hat{{\bf z}}\times\hat{{\bf k}}, \label{eq:eTE} \\
\hat{\boldsymbol{e}}_{\text{TM}}^{\pm}\left({\bf k}\right) & =\hat{\boldsymbol{e}}_{\text{TE}}^{\pm}\left({\bf k}\right)\times\hat{{\bf K}}^{\pm}.
\label{eq:eTM} 
\end{align}
Besides, the product $\left[\hat{\boldsymbol{e}}_{p^{\prime}}^{-}\left({\bf k}^{\prime}\right)\right]_{m}\left[\hat{\boldsymbol{e}}_{p}^{+}\left({\bf k}\right)\right]_{n}$ can be viewed as components of a $3 \times 3$ matrix, so that we have one matrix for each combination of field polarizations ($p^\prime$ and $p$).
Thus, using Eqs. \eqref{eq:eTE} and \eqref{eq:eTM}, one obtains that
\begin{widetext}
\begin{align}
\left[\hat{\boldsymbol{e}}_{\text{TE}}^{-}\left({\bf k}^{\prime}\right)\right]_{m}\left[\hat{\boldsymbol{e}}_{\text{TE}}^{+}\left({\bf k}\right)\right]_{n} & =\frac{1}{\left|{\bf k}^{\prime}\right|\left|{\bf k}\right|}\left(\begin{array}{ccc}
k_{y}^{\prime}k_{y} & -k_{y}^{\prime}k_{x} & 0\\
-k_{x}^{\prime}k_{y} & k_{x}^{\prime}k_{x} & 0\\
0 & 0 & 0
\end{array}\right); \\
\left[\hat{\boldsymbol{e}}_{\text{TE}}^{-}\left({\bf k}^{\prime}\right)\right]_{m}\left[\hat{\boldsymbol{e}}_{\text{TM}}^{+}\left({\bf k}\right)\right]_{n} & =\frac{1}{\left|{\bf k}^{\prime}\right|\left|{\bf k}\right|}\frac{c}{\xi}\left(\begin{array}{ccc}
-k_{y}^{\prime}\kappa k_{x} & -k_{y}^{\prime}\kappa k_{y} & -ik_{y}^{\prime}\left|{\bf k}\right|^{2}\\
k_{x}^{\prime}\kappa k_{x} & k_{x}^{\prime}\kappa k_{y} & ik_{x}^{\prime}\left|{\bf k}\right|^{2}\\
0 & 0 & 0
\end{array}\right); \\
\left[\hat{\boldsymbol{e}}_{\text{TM}}^{-}\left({\bf k}^{\prime}\right)\right]_{m}\left[\hat{\boldsymbol{e}}_{\text{TE}}^{+}\left({\bf k}\right)\right]_{n} & =\frac{1}{\left|{\bf k}^{\prime}\right|\left|{\bf k}\right|}\frac{c}{\xi}\left(\begin{array}{ccc}
\kappa^{\prime}k_{x}^{\prime}k_{y} & -\kappa^{\prime}k_{x}^{\prime}k_{x} & 0\\
\kappa^{\prime}k_{y}^{\prime}k_{y} & -\kappa^{\prime}k_{y}^{\prime}k_{x} & 0\\
-ik_{y}\left|{\bf k}^{\prime}\right|^{2} & ik_{x}\left|{\bf k}^{\prime}\right|^{2} & 0
\end{array}\right); \\
\left[\hat{\boldsymbol{e}}_{\text{TM}}^{-}\left({\bf k}^{\prime}\right)\right]_{m}\left[\hat{\boldsymbol{e}}_{\text{TM}}^{+}\left({\bf k}\right)\right]_{n} & =\frac{1}{\left|{\bf k}^{\prime}\right|\left|{\bf k}\right|}\frac{c^{2}}{\xi^{2}}\left(\begin{array}{ccc}
-\kappa^{\prime}k_{x}^{\prime}\kappa k_{x} & -\kappa^{\prime}k_{x}^{\prime}\kappa k_{y} & -i\kappa^{\prime}k_{x}^{\prime}\left|{\bf k}\right|^{2}\\
-\kappa^{\prime}k_{y}^{\prime}\kappa k_{x} & -\kappa^{\prime}k_{y}^{\prime}\kappa k_{y} & -i\kappa^{\prime}k_{y}^{\prime}\left|{\bf k}\right|^{2}\\
i\kappa k_{x}\left|{\bf k}^{\prime}\right|^{2} & i\kappa k_{y}\left|{\bf k}^{\prime}\right|^{2} & -\left|{\bf k}^{\prime}\right|^{2}\left|{\bf k}\right|^{2}
\end{array}\right).
\end{align}
\end{widetext}

Now that we found the reflection operator for an anisotropic polarizable particle, the interaction energy between it and a general surface is computed by substituting Eq. \eqref{eq:operador-RP} into Eq. \eqref{eq:energia-geral}, but taking $z_0 = 0$ in the former, because the propagation of the field along the $z$-axis is already taken into account in Eq. \eqref{eq:energia-geral} [described by the factor $\exp[-\left(\kappa+\kappa^{\prime}\right)z_{0}]$].
Thus, when performing this, one obtains
\begin{align}
U\left({\bf r}_{0}\right) & =\frac{\hbar}{\epsilon_{0}c^{2}}\int_{0}^{\infty}\frac{d\xi}{2\pi}\xi^{2}\sum_{m,n}\alpha_{mn}\left(i\xi\right)\nonumber \\
 & \times\int\frac{d^{2}{\bf k}}{\left(2\pi\right)^{2}}\int\frac{d^{2}{\bf k}^{\prime}}{\left(2\pi\right)^{2}}\frac{e^{-\left(\kappa+\kappa^{\prime}\right)z_{0}}}{2\kappa^{\prime}}e^{i\left({\bf k}-{\bf k}^{\prime}\right)\cdot{\bf r}_{0||}}\nonumber \\
 & \times\sum_{p,p^{\prime}}\left\langle {\bf k},p\right|{\cal R}_{S}\left|{\bf k}^{\prime},p^{\prime}\right\rangle \left[\hat{\boldsymbol{e}}_{p^{\prime}}^{-}\left({\bf k}^{\prime}\right)\right]_{m}\left[\hat{\boldsymbol{e}}_{p}^{+}\left({\bf k}\right)\right]_{n}.
\label{eq:energia-part-anisotr}
\end{align}
Note that, when we consider an isotropic particle [$\alpha_{mn}\left(i\xi\right) = \alpha\left(i\xi\right) \delta_{mn}$], we recover the expression for $U\left({\bf r}_{0}\right)$ obtained in Ref. \cite{Messina-PRA-2009} [Eq. (11) of this reference].

\subsection{The consideration of a corrugated surface}

As discussed in Ref. \cite{Messina-PRA-2009}, the calculation of the reflection operator for a corrugated surface is a highly nontrivial problem.
One way to do this is by writing this operator as a perturbative expansion in powers of $h(x,y)$.
Such calculation was already done in Refs. \cite{Neto-EPL-2005, Neto-PRA-2005} considering that the corrugation amplitude [$\text{max}|h(x,y)| = a$] is the smallest length scale in the problem.
The solution obtained from this calculation was used in Ref. \cite{Messina-PRA-2009} up to first perturbative order.
Following Ref. \cite{Messina-PRA-2009}, let us also consider the reflection operator for a corrugated surface as a perturbative expansion in powers of $h$.
Assuming that $a \ll z_0$, we can consider this perturbative expansion up to the first order, i.e.:
\begin{equation}
{\cal R}_{S}\approx{\cal R}_{S}^{(0)}+{\cal R}_{S}^{(1)},
\end{equation}
where ${\cal R}_{S}^{(0)} $ is the unperturbed solution which describes the specular reflection of the field on a plane surface, whereas ${\cal R}_{S}^{(1)}$ is the first-order correction to ${\cal R}_{S}^{(0)}$ due to the surface corrugation.
In this context, the Casimir-Polder interaction can be written as the sum
\begin{equation}
U\left({\bf r}_{0}\right) \approx U^{(0)}\left(z_{0}\right)+U^{(1)}\left({\bf r}_{0}\right).
\end{equation}
Note that the dependence of $U^{(1)}$ on the variables $x_0$ and $y_0$ gives rise to a lateral force (a force parallel to the reference plane $z=0$) that acts on the particle.
Since we are interested only in the behavior of this lateral force, we focus our attention only on $U^{(1)}$, which involves ${\cal R}_{S}^{(1)}$.

The matrix elements of the first-order reflection operator ${\cal R}_{S}^{(1)}$ can be written as \cite{Neto-PRA-2005, Messina-PRA-2009}
\begin{equation}
\langle{\bf k},p|{\cal R}_{S}^{(1)}|{\bf k}^{\prime},p^{\prime}\rangle=R_{pp^{\prime}}^{(1)}\left({\bf k},{\bf k}^{\prime}\right)H\left({\bf k}-{\bf k}^{\prime}\right),
\label{eq:operador-RS}
\end{equation}
where $H\left({\bf k}\right)$ is the Fourier transform of $h(x,y)$, and the components of the matrix $R_{pp^{\prime}}^{(1)}$ are given by \cite{Neto-PRA-2005, Messina-PRA-2009}:
\begin{align}
R_{\text{TE},\text{TE}}^{(1)}({\bf k},{\bf k}^{\prime}) & =2\kappa^{\prime}\frac{\kappa-\kappa_{t}}{\kappa^{\prime}+\kappa_{t}^{\prime}}C, \label{eq:R-TE-TE} \\
R_{\text{TE},\text{TM}}^{(1)}({\bf k},{\bf k}^{\prime}) & =2\kappa^{\prime}\frac{c}{\xi}\frac{\kappa_{t}^{\prime}\left(\kappa-\kappa_{t}\right)}{\epsilon\left(i\xi\right)\kappa^{\prime}+\kappa_{t}^{\prime}}S, \label{eq:R-TE-TM} \\
R_{\text{TM},\text{TE}}^{(1)}({\bf k},{\bf k}^{\prime}) & =2\kappa^{\prime}\frac{\xi/c}{\left(\kappa^{\prime}+\kappa_{t}^{\prime}\right)}\frac{\left[\epsilon\left(i\xi\right)\kappa-\kappa_{t}\right]\kappa_{t}S}{\frac{\xi^{2}}{c^{2}}-\kappa^{2}\left[\epsilon\left(i\xi\right)+1\right]}, \label{eq:R-TM-TE} \\
R_{\text{TM},\text{TM}}^{(1)}({\bf k},{\bf k}^{\prime}) & =-2\kappa^{\prime}\frac{\epsilon(i\xi)\kappa-\kappa_{t}}{\epsilon(i\xi)\kappa^{\prime}+\kappa_{t}^{\prime}}\frac{[\epsilon(i\xi)|{\bf k}||{\bf k}^{\prime}|+\kappa_{t}\kappa_{t}^{\prime}C]}{\frac{\xi^{2}}{c^{2}}-\kappa^{2}[\epsilon(i\xi)+1]},
\label{eq:R-TM-TM}
\end{align}
with $\kappa_{t}=\sqrt{\epsilon\left(i\xi\right) \xi^{2}/c^{2}+\left|{\bf k}\right|^{2}}$ and
\begin{align}
S & =\sin\left(\phi-\phi^{\prime}\right)=\frac{k_{y}k_{x}^{\prime}-k_{x}k_{y}^{\prime}}{\left|{\bf k}\right|\left|{\bf k}^{\prime}\right|},  \label{eq:Sin}  \\
C & =\cos\left(\phi-\phi^{\prime}\right)=\frac{k_{x}k_{x}^{\prime}+k_{y}k_{y}^{\prime}}{\left|{\bf k}\right|\left|{\bf k}^{\prime}\right|}.
\label{eq:Cos}
\end{align}
Thus, substituting Eq. \eqref{eq:operador-RS} into Eq. \eqref{eq:energia-part-anisotr}, we can write $U^{(1)}$ in a similar way to Ref. \cite{Messina-PRA-2009}, namely
\begin{equation}
U^{(1)}\left({\bf r}_{0}\right)=\int\frac{d^{2}{\bf k}}{\left(2\pi\right)^{2}}e^{i{\bf k}\cdot{\bf r}_{0 ||}}g\left({\bf k},z_{0}\right)H\left({\bf k}\right),
\label{eq:u1}
\end{equation}
where $g\left({\bf k},z_{0}\right)$ is the response function given by
\begin{equation}
g({\bf k},z_{0})=\frac{\hbar}{\epsilon_{0}c^{2}}\int_{0}^{\infty}\frac{d\xi}{2\pi}\xi^{2}\sum_{m,n}\alpha_{mn}(i\xi)\int\frac{d^{2}{\bf k}^{\prime}}{(2\pi)^{2}}a_{{\bf k}^{\prime},{\bf k}^{\prime}-{\bf k}}^{mn},
\label{eq:funcao-g}
\end{equation}
with 
\begin{align}
a_{{\bf k}^{\prime},{\bf k}^{\prime\prime}}^{mn} & =\frac{e^{-(\kappa^{\prime}+\kappa^{\prime\prime})z_{0}}}{2\kappa^{\prime\prime}}\nonumber \\
 & \times\sum_{p^{\prime},p^{\prime\prime}}R_{p^{\prime},p^{\prime\prime}}^{(1)}({\bf k}^{\prime},{\bf k}^{\prime\prime})[\hat{e}_{p^{\prime\prime}}^{-}({\bf k}^{\prime\prime})]_{m}[\hat{e}_{p^{\prime}}^{+}({\bf k}^{\prime})]_{n}.
\label{eq:funcao-amn}
\end{align}
We remark that these results are a generalization of those found in Refs. \cite{Nogueira-PRA-2021, Queiroz-PRA-2021} since they take into account the realistic properties for the surface and retardation effects in the interaction.
Besides this, they also generalize Eqs. (16) and (17) of Ref. \cite{Messina-PRA-2009} by the consideration of an anisotropic electrically polarizable particle.
Using these results, we later investigate the behavior of $U^{(1)}$ to obtain information about the lateral force acting on the particle.
But before this, let us discuss these results within the vdW and CP regimes.

\subsection{Van der Waals regime}

The vdW regime is the limiting case where retardation effects can be neglected, which occurs when $z_{0} \ll \lambda_{P},\lambda_{S}$, with $\lambda_{P}$ and $\lambda_{S}$ being typical wavelengths that characterize the optical responses of the particle and the surface, respectively.
In this regime, it is assumed that the particle-surface interaction is instantaneous, which means that the limit $c \to \infty$ can be considered.
Thus, by performing $c \to \infty$ in Eqs. \eqref{eq:u1}-\eqref{eq:funcao-amn}, one obtains
\begin{equation}
U^{(1)}_{\text{vdW}}\left({\bf r}_{0}\right)=\int\frac{d^{2}{\bf k}}{\left(2\pi\right)^{2}}e^{i{\bf k}\cdot{\bf r}_{0 ||}}g_{\text{vdW}}\left({\bf k},z_{0}\right)H\left({\bf k}\right),
\label{eq:u1-vdW}
\end{equation}
where
\begin{align}
g_{\text{vdW}}({\bf k},z_{0}) & =\frac{\hbar}{\epsilon_{0}}\sum_{m,n}\int_{0}^{\infty}\frac{d\xi}{2\pi}\alpha_{mn}(i\xi)\int\frac{d^{2}{\bf k}^{\prime}}{(2\pi)^{2}}e^{-(|{\bf k}^{\prime}|+|{\bf k}^{\prime\prime}|)z_{0}}\nonumber \\
 & \times\frac{[\epsilon(i\xi)-1]}{[\epsilon(i\xi)+1]^{2}}\left[\epsilon(i\xi)+\frac{k_{x}^{\prime}k_{x}^{\prime\prime}+k_{y}^{\prime}k_{y}^{\prime\prime}}{|{\bf k}^{\prime}||{\bf k}^{\prime\prime}|}\right]G_{mn},
\label{eq:funcao-g-vdW}
\end{align}
with ${\bf k}^{\prime\prime}={\bf k}^{\prime}-{\bf k}$, and
\begin{equation}
G_{mn}=\left(\begin{array}{ccc}
-k_{x}^{\prime\prime}k_{x}^{\prime} & -k_{x}^{\prime\prime}k_{y}^{\prime} & -ik_{x}^{\prime\prime}|{\bf k}^{\prime}|\\
-k_{y}^{\prime\prime}k_{x}^{\prime} & -k_{y}^{\prime\prime}k_{y}^{\prime} & -ik_{y}^{\prime\prime}|{\bf k}^{\prime}|\\
ik_{x}^{\prime}|{\bf k}^{\prime\prime}| & ik_{y}^{\prime}|{\bf k}^{\prime\prime}| & -|{\bf k}^{\prime\prime}||{\bf k}^{\prime}|
\end{array}\right). 
\label{eq:Amn}
\end{equation}
Such result provides the vdW interaction between an anisotropic polarizable particle and a corrugated surface made of a real dispersive material.

When considering a nondispersive surface, one has $\epsilon(i\xi)=\epsilon$, so that the integration over $\xi$ has to be performed only on the polarizability tensor components $\alpha_{mn}(i\xi)$.
One can write these components as \cite{Buhmann-DispersionForces-II}
\begin{equation}
\alpha_{mn}\left(i\xi\right)=\frac{2}{\hbar}\sum_{l\neq k}\frac{\omega_{lk}\langle k|\hat{d}_{m}|l\rangle\langle l|\hat{d}_{n}|k\rangle}{\omega_{lk}^{2}+\xi^{2}},
\label{eq:polarizabilidade-geral}
\end{equation}
where $\hat{d}_i (i=x,y,z)$ are the components of the dipole moment operator and $\omega_{lk}$ is the frequency of a typical transition between the states $l$ and $k$ of the particle.
Thus, by performing the $\xi$-integration on \eqref{eq:polarizabilidade-geral}, one obtains
\begin{equation}
\int_{0}^{\infty}d\xi\alpha_{mn}\left(i\xi\right)=\frac{\pi}{\hbar}\langle\hat{d}_{m}\hat{d}_{n}\rangle.
\end{equation}
By substituting this result into Eq. \eqref{eq:funcao-g-vdW}, and performing the integrals over ${\bf k}^{\prime}$, Eqs. \eqref{eq:u1-vdW}-\eqref{eq:Amn} recover the result for $U^{(1)}_{\text{vdW}}$ obtained in Ref. \cite{Queiroz-PRA-2021} [Eq. (50) of this reference with $\epsilon_{2}=1$].
In addition, when considering a perfectly conducting surface [$\epsilon(i\xi) \to \infty$], Eqs. \eqref{eq:u1-vdW}-\eqref{eq:Amn} recover the result for $U^{(1)}_{\text{vdW}}$ obtained in Ref. \cite{Nogueira-PRA-2021} [Eq. (14) of this reference].

\subsection{Casimir-Polder regime}

The CP regime is the limiting case where retardation effects must be taken into account, which occurs when $z_{0} \gg \lambda_{P},\lambda_{S}$.
In this regime, $U^{(1)}_{\text{CP}}$ is given by Eqs. \eqref{eq:u1}-\eqref{eq:funcao-amn} with $\alpha_{mn}\left(i\xi\right) \to \alpha_{mn}\left(0\right)$ and $\epsilon\left(i\xi\right) \to \epsilon\left(0\right)$.
When considering metallic surfaces, one can also perform the limit $\epsilon\left(0\right) \to \infty$ in Eqs. \eqref{eq:u1}-\eqref{eq:funcao-amn}, which acts only on the components of the matrix $R_{pp^{\prime}}^{(1)}$ [Eqs. \eqref{eq:R-TE-TE}-\eqref{eq:R-TM-TM}] leading to \cite{Dalvit-JPA-2008}
\begin{align}
\lim_{\epsilon\to\infty}R_{\text{TE},\text{TE}}^{(1)}({\bf k},{\bf k}^{\prime}) & =-2\kappa^{\prime}C, \\
\lim_{\epsilon\to\infty}R_{\text{TE},\text{TM}}^{(1)}({\bf k},{\bf k}^{\prime}) & =-2\frac{\xi}{c}S, \\
\lim_{\epsilon\to\infty}R_{\text{TM},\text{TE}}^{(1)}({\bf k},{\bf k}^{\prime}) & =-2\frac{\xi}{c}\frac{\kappa^{\prime}}{\kappa}S, \\
\lim_{\epsilon\to\infty}R_{\text{TM},\text{TM}}^{(1)}({\bf k},{\bf k}^{\prime}) & =\frac{2}{\kappa}\left(\left|{\bf k}\right|\left|{\bf k}^{\prime}\right|+\frac{\xi^{2}}{c^{2}}C\right),
\end{align}
with $S$ and $C$ given by Eqs. \eqref{eq:Sin} and \eqref{eq:Cos}.
%

\section{Sinusoidal corrugation}
\label{sec-sinusoidal}

\begin{figure}[h]
\centering 
\epsfig{file=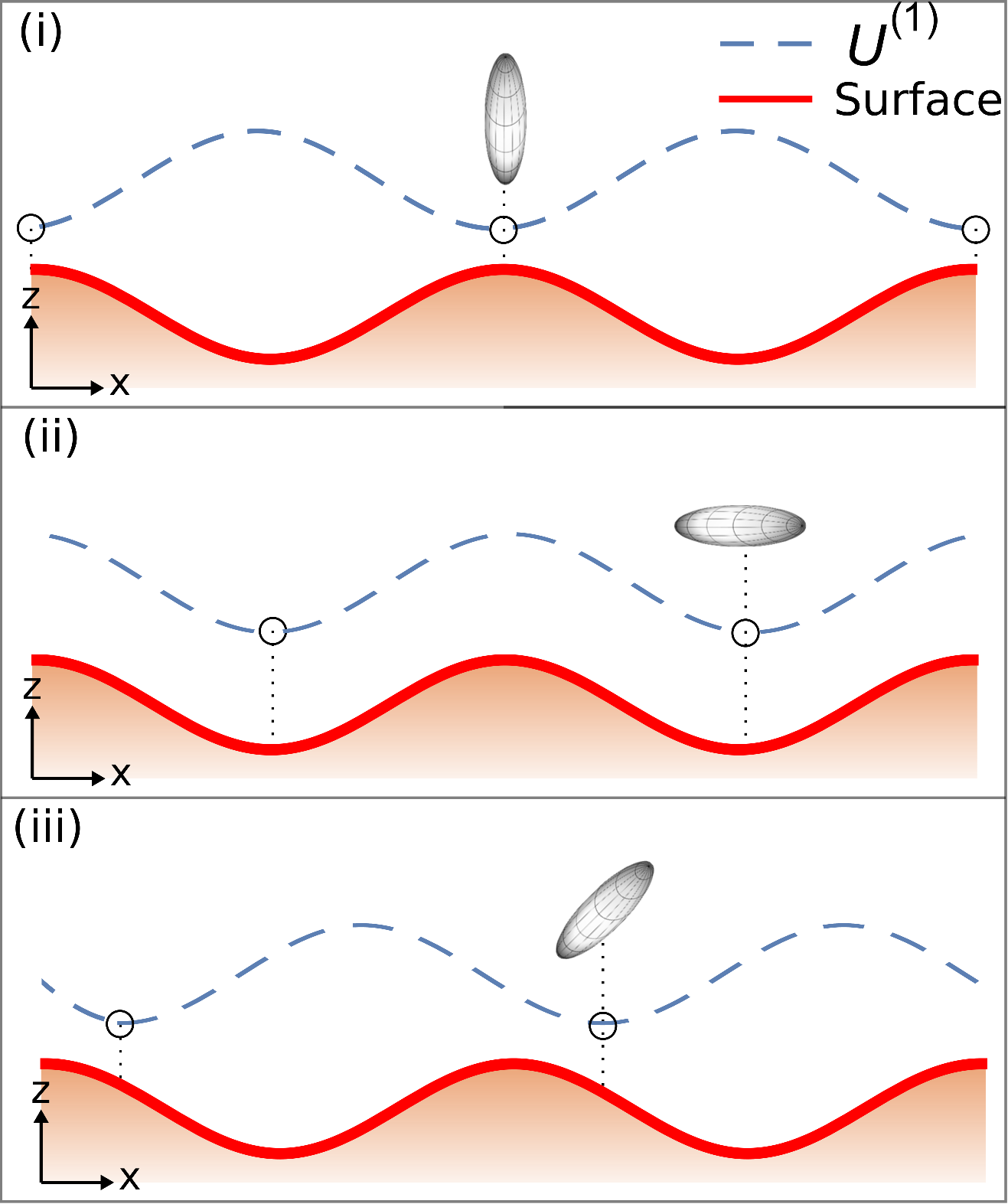,  width=0.6 \linewidth}  
\caption{
Illustration of a neutral anisotropic particle (represented by an ellipsoid) interacting with a corrugated surface (solid lines).
The stable equilibrium points (indicated by the circles) of $U^{(1)}(x_{0},z_{0})$ (dashed lines), with respect to $x_0$, can be over the corrugation peaks [peak regime, (i)], valleys [valley regime, (ii)], or over intermediate points between a peak and a valley [intermediate regime, (iii)]. 
}
\label{fig:regimes-geral}
\end{figure}

To investigate the behavior of $U^{(1)}$, it is convenient to choose a corrugation profile for the surface.
Let us investigate the case of a sinusoidal corrugated surface with amplitude $a$ and corrugation period $\lambda_c$, which is described by $h(x,y)= a \cos(k_c x)$, where $k_c = 2\pi/\lambda_c$ and $a\ll z_0$.
For this case, substituting the Fourier transform of $h$ into Eq. \eqref{eq:u1}, and performing the $\textbf{k}$-integration, one obtains
\begin{equation}
U^{(1)}(x_{0},z_{0})=\frac{a}{2}[e^{ik_{c}x_{0}}g(\tilde{{\bf k}},z_{0})+e^{-ik_{c}x_{0}}g(-\tilde{{\bf k}},z_{0})],
\end{equation}
where $\tilde{{\bf k}} = k_c \hat{{\bf x}}$ and $g$ is given by Eqs. \eqref{eq:funcao-g} and \eqref{eq:funcao-amn}.
By carefully inspecting the ${\bf k}^{\prime}$-integration $\int d^{2}{\bf k}^{\prime} a_{{\bf k}^{\prime},{\bf k}^{\prime}+\tilde{{\bf k}}}^{mn}$ within $g(-\tilde{{\bf k}},z_{0})$, we find that
\begin{align}
\int d^{2}{\bf k}^{\prime} & a_{{\bf k}^{\prime},{\bf k}^{\prime}+\tilde{{\bf k}}}^{mn}=\nonumber \\
 & \int d^{2}{\bf k}^{\prime}\left(\begin{array}{ccc}
a_{{\bf k}^{\prime},{\bf k}^{\prime}-\tilde{{\bf k}}}^{xx} & 0 & -a_{{\bf k}^{\prime},{\bf k}^{\prime}-\tilde{{\bf k}}}^{xz}\\
0 & a_{{\bf k}^{\prime},{\bf k}^{\prime}-\tilde{{\bf k}}}^{yy} & 0\\
-a_{{\bf k}^{\prime},{\bf k}^{\prime}-\tilde{{\bf k}}}^{xz} & 0 & a_{{\bf k}^{\prime},{\bf k}^{\prime}-\tilde{{\bf k}}}^{zz}
\end{array}\right),
\end{align}
so that $U^{(1)}(x_{0},z_{0})$ is written as
\begin{align}
U^{(1)} (x_{0},z_{0}) & =\frac{a\hbar}{8\epsilon_{0}\pi^{3}}\int_{0}^{\infty}d\xi\int d^{2}{\bf k}^{\prime}\frac{\xi^{2}}{c^{2}}\nonumber \\
 & \times[\sum_{m}\alpha_{mm}(i\xi)a_{{\bf k}^{\prime},{\bf k}^{\prime}-\tilde{{\bf k}}}^{mm}\cos(k_{c}x_{0})\nonumber \\
 & +2i\alpha_{xz}(i\xi)a_{{\bf k}^{\prime},{\bf k}^{\prime}-\tilde{{\bf k}}}^{xz}\sin(k_{c}x_{0})]. 
\label{eq:u1-sin-cos}
\end{align}
By defining
\begin{align}
{\cal V}_{xx} (k_{c},z_{0})& =\int_{0}^{\infty}d\xi\int d^{2}{\bf k}^{\prime}\frac{\xi^{2}}{c^{2}}\alpha_{xx}\left(i\xi\right)a_{{\bf k}^{\prime},{\bf k}^{\prime}-\tilde{{\bf k}}}^{xx},  \label{eq:vxx} \\
{\cal V}_{yy} (k_{c},z_{0})& =\int_{0}^{\infty}d\xi\int d^{2}{\bf k}^{\prime}\frac{\xi^{2}}{c^{2}}\alpha_{yy}\left(i\xi\right)a_{{\bf k}^{\prime},{\bf k}^{\prime}-\tilde{{\bf k}}}^{yy},  \label{eq:vyy} \\
{\cal V}_{zz} (k_{c},z_{0})& =\int_{0}^{\infty}d\xi\int d^{2}{\bf k}^{\prime}\frac{\xi^{2}}{c^{2}}\alpha_{zz}\left(i\xi\right)a_{{\bf k}^{\prime},{\bf k}^{\prime}-\tilde{{\bf k}}}^{zz},  \label{eq:vzz} \\
{\cal V}_{xz} (k_{c},z_{0})& =2i\int_{0}^{\infty}d\xi\int d^{2}{\bf k}^{\prime}\frac{\xi^{2}}{c^{2}}\alpha_{xz}\left(i\xi\right)a_{{\bf k}^{\prime},{\bf k}^{\prime}-\tilde{{\bf k}}}^{xz},
\label{eq:vxz}
\end{align}
we can write Eq. \eqref{eq:u1-sin-cos} as
\begin{equation}
U^{(1)}(x_{0},z_{0})=\frac{a\hbar}{8\pi^{3}\epsilon_{0}}A(k_{c},z_{0})\cos[k_{c}x_{0}-\delta(k_{c},z_{0})],
\label{eq:u1-senoidal}
\end{equation}
where
\begin{equation}
A(k_{c},z_{0})=\sqrt{{\cal V}_{\text{Sum}}^{2}+{\cal V}_{xz}^{2}},
\label{eq:A}
\end{equation}
with ${\cal V}_{\text{Sum}}={\cal V}_{xx}+{\cal V}_{yy}+{\cal V}_{zz}$, and $\delta(k_{c},z_{0})$ is a nontrivial phase function defined by
\begin{equation}
\delta(k_{c},z_{0})=\arctan\left(\frac{{\cal V}_{xz}}{{\cal V}_{\text{Sum}}}\right).
\label{eq:funcao-delta}
\end{equation}
From Eq. \eqref{eq:u1-senoidal}, considering the behavior of $U^{(1)}$ with respect to $x_0$, one can see that the stable equilibrium points of $U^{(1)}$ can be over the corrugation peaks $(\delta = \pi)$, valleys $(\delta = 0)$ or over intermediate points between a peak and a valley $(\delta \neq 0, \pi)$.
Such possibilities were first investigated in Refs. \cite{Nogueira-PRA-2021, Queiroz-PRA-2021} and were called peak, valley and intermediate regimes, respectively (see Fig. \ref{fig:regimes-geral}).
In both Refs. \cite{Nogueira-PRA-2021, Queiroz-PRA-2021}, these effects were studied within the vdW regime of the interaction, but the corrugated surface was considered as a perfect conductor in Ref. \cite{Nogueira-PRA-2021}, while it was considered as a nondispersive surface in Ref. \cite{Queiroz-PRA-2021}.
Thus, Eqs. \eqref{eq:u1-senoidal}-\eqref{eq:funcao-delta} generalize the results obtained in these references by considering both realistic material properties for the surface and retardation effects in the interaction, enabling us to investigate their effects on the occurrence of the regimes of the lateral force.
%

\subsection{Van der Waals regime of Eq. \eqref{eq:u1-senoidal}}

For our purposes, it is interesting to calculate Eq. \eqref{eq:u1-senoidal} in the vdW regime.
When performing the limit $c \to \infty$ on this equation, one can see that it acts only on the functions ${\cal V}_{mn}$, enabling us to calculate the integrals over ${\bf k}^{\prime}$, obtaining:
\begin{widetext}
\begin{align}
{\cal V}_{xx}^{\text{vdW}} & =-\frac{3\pi}{64z_{A}^{4}}\int_{0}^{\infty}d\xi\alpha_{xx}\left(i\xi\right)\frac{\left[\epsilon\left(i\xi\right)-1\right]}{\left[\epsilon\left(i\xi\right)+1\right]^{2}}\left[\epsilon\left(i\xi\right){\cal K}_{xx}^{\text{cond}}\left(k_{c}z_{0}\right)+{\cal K}_{xx}^{\text{diel}}\left(k_{c}z_{0}\right)\right],  \label{eq:vxx-vdw} \\
{\cal V}_{yy}^{\text{vdW}} & =-\frac{3\pi}{64z_{A}^{4}}\int_{0}^{\infty}d\xi\alpha_{yy}\left(i\xi\right)\frac{\left[\epsilon\left(i\xi\right)-1\right]}{\left[\epsilon\left(i\xi\right)+1\right]^{2}}\left[\epsilon\left(i\xi\right){\cal K}_{yy}^{\text{cond}}\left(k_{c}z_{0}\right)+{\cal K}_{yy}^{\text{diel}}\left(k_{c}z_{0}\right)\right],  \label{eq:vyy-vdw} \\
{\cal V}_{zz}^{\text{vdW}} & =-\frac{3\pi}{64z_{A}^{4}}\int_{0}^{\infty}d\xi\alpha_{zz}\left(i\xi\right)\frac{\left[\epsilon\left(i\xi\right)-1\right]}{\left[\epsilon\left(i\xi\right)+1\right]^{2}}\left[\epsilon\left(i\xi\right){\cal K}_{zz}^{\text{cond}}\left(k_{c}z_{0}\right)+{\cal K}_{zz}^{\text{diel}}\left(k_{c}z_{0}\right)\right],  \label{eq:vzz-vdw} \\
{\cal V}_{xz}^{\text{vdW}} & =\frac{3\pi}{32z_{A}^{4}}\int_{0}^{\infty}d\xi\alpha_{xz}\left(i\xi\right)\frac{\left[\epsilon\left(i\xi\right)-1\right]}{\left[\epsilon\left(i\xi\right)+1\right]^{2}}\left[\epsilon\left(i\xi\right){\cal K}_{xz}^{\text{cond}}\left(k_{c}z_{0}\right)+{\cal K}_{xz}^{\text{diel}}\left(k_{c}z_{0}\right)\right],
\label{eq:vxz-vdw} 
\end{align}
where the functions $ {\cal K}_{mn}^{\text{cond}} $ and $ {\cal K}_{mn}^{\text{diel}} $ are given by \cite{Queiroz-PRA-2021}:
\begin{align}
\mathcal{K}_{xx}^{\text{cond}}\left(u\right) & =u^{3}K_{3}\left(u\right)-u^{4}K_{2}\left(u\right), & \mathcal{K}_{xx}^{\text{diel}}\left(u\right) & =\left[\frac{56}{3}u^{2}+u^{4}\right]K_{2}\left(u\right)-\frac{11}{3}u^{3}K_{3}\left(u\right),  \label{eq:kxx-vdw} \\
\mathcal{K}_{yy}^{\text{cond}}\left(u\right) & =u^{3}K_{3}\left(u\right), & \mathcal{K}_{yy}^{\text{diel}}\left(u\right) & =8u^{2}K_{2}\left(u\right)-u^{3}K_{3}\left(u\right),  \label{eq:kyy-vdw} \\
\mathcal{K}_{zz}^{\text{cond}}\left(u\right) & =\left[\frac{16}{3}u^{2}+u^{4}\right]K_{2}\left(u\right)+\frac{2}{3}u^{3}K_{3}\left(u\right), & \mathcal{K}_{zz}^{\text{diel}}\left(u\right) & =2u^{3}K_{3}\left(u\right)-u^{4}K_{2}\left(u\right),  \label{eq:kzz-vdw} \\
\mathcal{K}_{xz}^{\text{cond}}\left(u\right) & =\frac{8}{3}u^{3}K_{2}\left(u\right)-u^{4}K_{3}\left(u\right), & \mathcal{K}_{xz}^{\text{diel}}\left(u\right) & =u^{4}K_{3}\left(u\right)-\frac{16}{3}u^{3}K_{2}\left(u\right),
\label{eq:kxz-vdw} 
\end{align}
with $K_2$ and $K_3$ being modified Bessel functions of the second kind.
Thus, the vdW interaction between an anisotropic polarizable particle and a dispersive sinusoidal surface is given by Eqs. \eqref{eq:u1-senoidal}-\eqref{eq:funcao-delta} with $ {\cal V}_{mn} \to {\cal V}_{mn}^{\text{vdW}} $ given by Eqs. \eqref{eq:vxx-vdw}-\eqref{eq:kxz-vdw}.
We remark that depending on the models for $ \alpha_{mn}\left(i\xi\right) $ and $ \epsilon\left(i\xi\right) $ the $ \xi $-integrals in Eqs. \eqref{eq:vxx-vdw}-\eqref{eq:vxz-vdw} can also be performed.
Lastly, Eqs. \eqref{eq:vxx-vdw}-\eqref{eq:kxz-vdw} could also be obtained from Eqs. \eqref{eq:u1-vdW}-\eqref{eq:Amn} by considering in them a sinusoidal corrugation and performing the ${\bf k}^{\prime}$-integration.
\end{widetext}

\subsection{Model for $ \alpha_{mn}\left(i\xi\right) $}

Let us describe, for simplicity (but without loss of generality), the anisotropic polarizable particle as a prolate spheroidal nanoparticle.
This particle has a rotational symmetry axis (named particle axis), which we consider initially oriented in $z$-direction, such that its polarizability tensor is represented by the matrix 
\begin{equation}
\alpha_{mn}(i\xi) = \left(\begin{array}{ccc}
\alpha_{\perp}(i\xi) & 0 & 0\\
0 & \alpha_{\perp}(i\xi) & 0\\
0 & 0 & \alpha_{\parallel}(i\xi)
\end{array}\right),
\label{eq:alpha-diag}
\end{equation}
with \cite{Bimonte-PRD-2015}
\begin{align}
\alpha_{\parallel}(i\xi) & =\epsilon_{0}V\frac{\epsilon_{P}(i\xi)-1}{1+[\epsilon_{P}(i\xi)-1]d},  \label{eq:alpha-p} \\
\alpha_{\perp}(i\xi) & =\epsilon_{0}V\frac{\epsilon_{P}(i\xi)-1}{1+\frac{1}{2}[\epsilon_{P}(i\xi)-1](1-d)}, 
\label{eq:alpha-n}
\end{align}
which are the polarizabilities in the directions parallel and normal to the particle axis, respectively.
In these equations, $V$ is the spheroid's volume, $\epsilon_{P}$ is the dielectric permittivity of the material in which the nanoparticle is made of, and $d$ is the depolarizing factor of the spheroid which is written in terms of the particle aspect ratio $r$ (ratio between its length and width) as \cite{Bimonte-PRD-2015}
\begin{equation}
d=\frac{1}{1-r^{2}}+\frac{r\log\left(\frac{r+\sqrt{r^{2}-1}}{r-\sqrt{r^{2}-1}}\right)}{2\left(r^{2}-1\right)^{3/2}}. 
\label{eq:fator-depolarizacao}
\end{equation}
For a prolate spheroid, one has $r > 1$, so that one can note that the limit $r \to 1$ corresponds to the case where the particle is a sphere and its polarizability is isotropic, on the other hand, in the limit $r \to \infty$ the particle is like a needle and it is highly anisotropic.
For a generic orientation of the particle, the polarizability tensor is represented by a non-diagonal matrix and we can write its components in terms of the spherical angles $(\theta,\phi)$ as:
\begin{align}
\alpha_{xx} & =\alpha_{\perp}+\left(\alpha_{\parallel}-\alpha_{\perp}\right)\sin^{2}\theta\cos^{2}\phi,  \label{eq:alpha-xx}  \\
\alpha_{yy} & =\alpha_{\perp}+\left(\alpha_{\parallel}-\alpha_{\perp}\right)\sin^{2}\theta\sin^{2}\phi,  \label{eq:alpha-yy}  \\
\alpha_{zz} & =\alpha_{\perp}+\left(\alpha_{\parallel}-\alpha_{\perp}\right)\cos^{2}\theta,  \label{eq:alpha-zz}  \\
\alpha_{xy} & =\frac{\alpha_{\parallel}-\alpha_{\perp}}{2}\sin2\phi\sin^{2}\theta,  \label{eq:alpha-xy}  \\
\alpha_{xz} & =\frac{\alpha_{\parallel}-\alpha_{\perp}}{2}\sin2\theta\cos\phi,  \label{eq:alpha-xz}  \\
\alpha_{yz} & =\frac{\alpha_{\parallel}-\alpha_{\perp}}{2}\sin2\theta\sin\phi.
\label{eq:alpha-yz}
\end{align}
We remark that the polarizability tensor is a symmetric tensor, thus one has $\alpha_{yx}=\alpha_{xy}$, $\alpha_{zx}=\alpha_{xz}$ and $\alpha_{zy}=\alpha_{yz}$.

\subsection{Models for $ \epsilon\left(i\xi\right) $ and $ \epsilon_{P}\left(i\xi\right) $}

Let us consider our formulas for the case of a gold nanoparticle interacting with a corrugated gold surface, whose permittivities are both described by the plasma model as
\begin{equation}
\epsilon\left(i\xi\right)=\epsilon_{P}\left(i\xi\right)=1+\frac{\omega_{p}^{2}}{\xi^{2}},
\label{eq:plasma-model}
\end{equation}
where $\omega_p$ is the plasma frequency of the metal.
We can also write the plasma frequency in terms of the plasma wavelength $\lambda_p$ as $\omega_p = 2\pi c/\lambda_p$.
For gold, we have $\omega_p \approx 1.385 \times 10^{16} \text{ rad/s}$, which results in $\lambda_p \approx 136 \text{ nm}$.
Since both the nanoparticle and the surface are considered to be made of gold, the value for $\lambda_p$ is the reference to determine if we are within the vdW or CP regimes, which are obtained when $z_0 \ll \lambda_p$ and $z_0 \gg \lambda_p$, respectively.
In addition, for the chosen situation, the $\xi$-integrals in Eqs. \eqref{eq:vxx-vdw}-\eqref{eq:vxz-vdw} can be performed, which means that we have an analytic expression for the vdW interaction between a gold spheroidal nanoparticle and a gold corrugated surface.

\section{Discussions}
\label{sec-discussions}

To investigate the behavior of the minimum points of $U^{(1)}$, one has to study the behavior of the function $\delta$, which depends on the functions ${\cal V}_{mn}$, as one can see in Eq. \eqref{eq:funcao-delta}.
Let us start making some general discussions on the behavior of the function $\delta$, which are similar to those made in Refs. \cite{Nogueira-PRA-2021, Queiroz-PRA-2021}.
From this equation, we can see that the occurrence of intermediate regimes just requires ${\cal V}_{xz} \neq 0$, which occurs when $\alpha_{xz} \neq 0$.
For the considered spheroidal particle, this happens when it is oriented such that $\sin2\theta\cos\phi \neq 0$ in Eq. \eqref{eq:alpha-xz}.
When this is not the case (for instance, when the particle is oriented such that its axis coincides with $x$, $y$ or $z$), one has $\alpha_{xz} = 0$ (thus, ${\cal V}_{xz} = 0$), and we have that the function $\delta$ can only assume the values $0$ or $\pi$, which correspond to valley and peak regimes, respectively.
In addition, we have one or the other of these regimes depending on the sign of ${\cal V}_{\text{Sum}}$, where we have valley regime when it is positive, and peak regime when negative [see Eq. \eqref{eq:funcao-delta}].
Among the functions ${\cal V}_{xx}$, ${\cal V}_{yy}$ and ${\cal V}_{zz}$, the only one that changes its sign is ${\cal V}_{xx}$, and thus it is responsible for the change of sign of ${\cal V}_{\text{Sum}}$.
As a consequence, considering Eqs. \eqref{eq:vxx}-\eqref{eq:vzz}, the greater $\alpha_{xx}$ is in relation to $\alpha_{yy}$ and $\alpha_{zz}$, the greater can be the occurrence of valley regime.
In other words, the more anisotropic the particle is in the direction of the corrugation, the greater can be the occurrence of valley regime.
Next, we investigate how the consideration of realistic material properties for the surface and of retardation in the interaction affects such occurrence.

\begin{figure}[h]
\centering
\epsfig{file=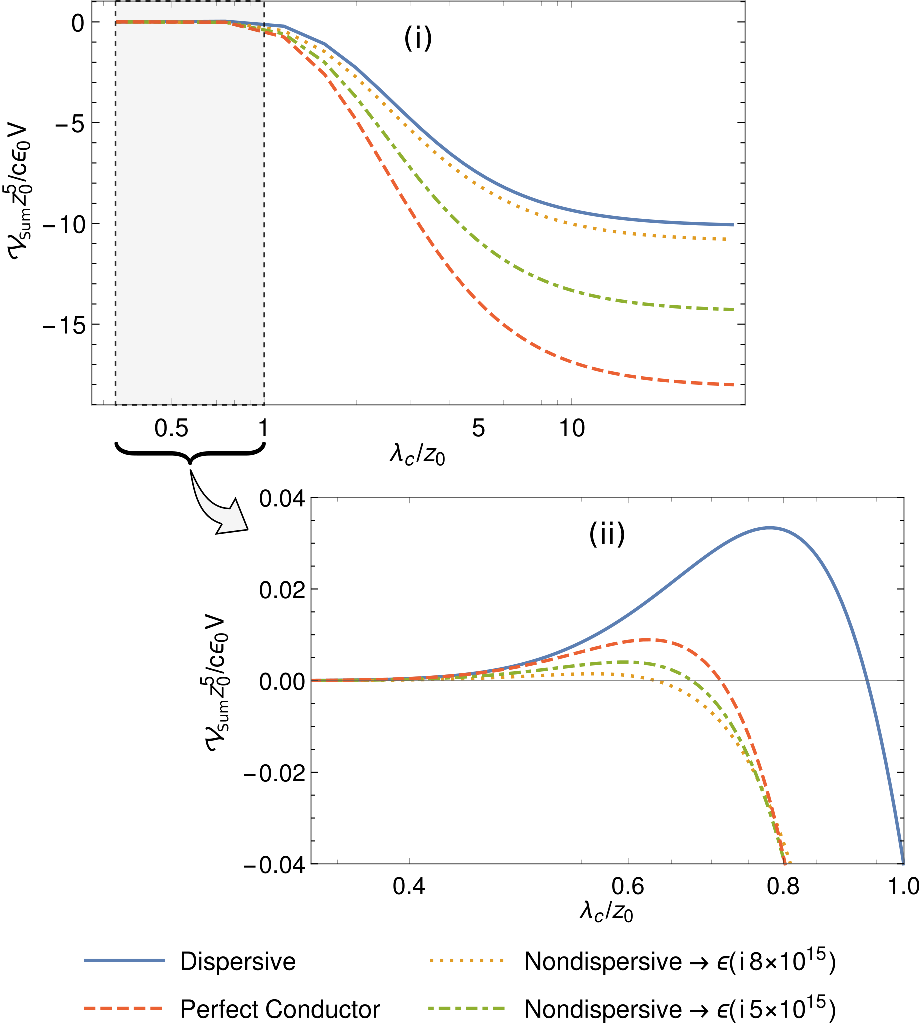, width=0.9 \linewidth}
\caption{
Behavior of ${\cal V}_{\text{Sum}}$ versus $\lambda_c/z_0$  for a gold spheroidal nanoparticle interacting with a gold sinusoidal surface within the vdW regime [Eqs. \eqref{eq:vxx-vdw}-\eqref{eq:vzz-vdw}].
The particle is oriented such that its axis is parallel to the $x$-axis, and $\alpha_{mn}$ is described by Eqs. \eqref{eq:alpha-xx}-\eqref{eq:alpha-yz} with $\theta=\pi/2$ and $\phi=0$.
The highlighted interval in panel (i) is shown in panel (ii) in a zoomed in view.
In each panel it is considered $z_0 = 30 \text{ nm}$ and $r = 2$, and it is shown the cases where the surface is considered dispersive (solid line), perfectly conductive (dashed line) and nondispersive [dotted line for $\epsilon(i 8 \times 10^{15})$ and dot-dashed for $\epsilon(i 5 \times 10^{15})$].
We remark that we have valley regime when ${\cal V}_{\text{Sum}} > 0$, and peak regime when ${\cal V}_{\text{Sum}} < 0$.
}
\label{fig:efeito-dispersao-vdW}
\end{figure}

Let us start considering the particle oriented such that its axis is parallel to the $x$-axis.
In this situation, $\alpha_{mn}$ is described by Eqs. \eqref{eq:alpha-xx}-\eqref{eq:alpha-yz} with $\theta=\pi/2$ and $\phi=0$, and, as discussed above, we have valley regime when ${\cal V}_{\text{Sum}} > 0$, or peak regime when ${\cal V}_{\text{Sum}} < 0$.
We begin investigating the effects of the consideration of realistic material properties for the surface on the occurrence of these regimes in the short distance limit (vdW regime).
For this, in Fig. \ref{fig:efeito-dispersao-vdW}, we consider Eqs. \eqref{eq:vxx-vdw}-\eqref{eq:vzz-vdw} and we show the behavior of ${\cal V}_{\text{Sum}}$, in terms of $\lambda_c/z_0$, for the cases in which the corrugated surface is considered dispersive, nondispersive [$\epsilon(i\xi)=\text{cte}$] and perfectly conductive [$\epsilon(i\xi) \to \infty$].
In this figure, it can be seen that the general behavior of ${\cal V}_{\text{Sum}}$ is similar for all cases, differing only numerically (this conclusion was first mentioned in Ref. \cite{Nogueira-PRA-2021} through rough estimates).
In Fig. \ref{fig:efeito-dispersao-vdW}(i), one notes that when the surface is considered as a perfect conductor one has the greatest magnitudes of ${\cal V}_{\text{Sum}}$, since its reflectivity is perfect (for the nondispersive cases, the surface has transparency and the mentioned magnitudes decrease, as discussed in Ref. \cite{Queiroz-PRA-2021}).
The opposite happens when the surface is dispersive, since in this case its reflectivity depends on the frequency of the field fluctuation, decreasing as the frequency increases.
Despite this, ${\cal V}_{\text{Sum}}$ can change its sign, as shown in Fig. \ref{fig:efeito-dispersao-vdW}(ii), which means that a transition between peak and valley regimes can occur.
Moreover, the consideration of realistic material properties for the surface can amplify the occurrence of valley regime, since the mentioned change of sign for the dispersive case occurs at a greater value of $\lambda_c/z_0$ than the other cases [as shown, for instance, in Fig. \ref{fig:efeito-dispersao-vdW}(ii)].
In other words, although the consideration of realistic material properties for the surface weakens the interaction, it can increase the range of $\lambda_c/z_0$ values in which the valley regime occurs, contributing to a greater occurrence of this regime.

\begin{figure}[h]
\centering
\epsfig{file=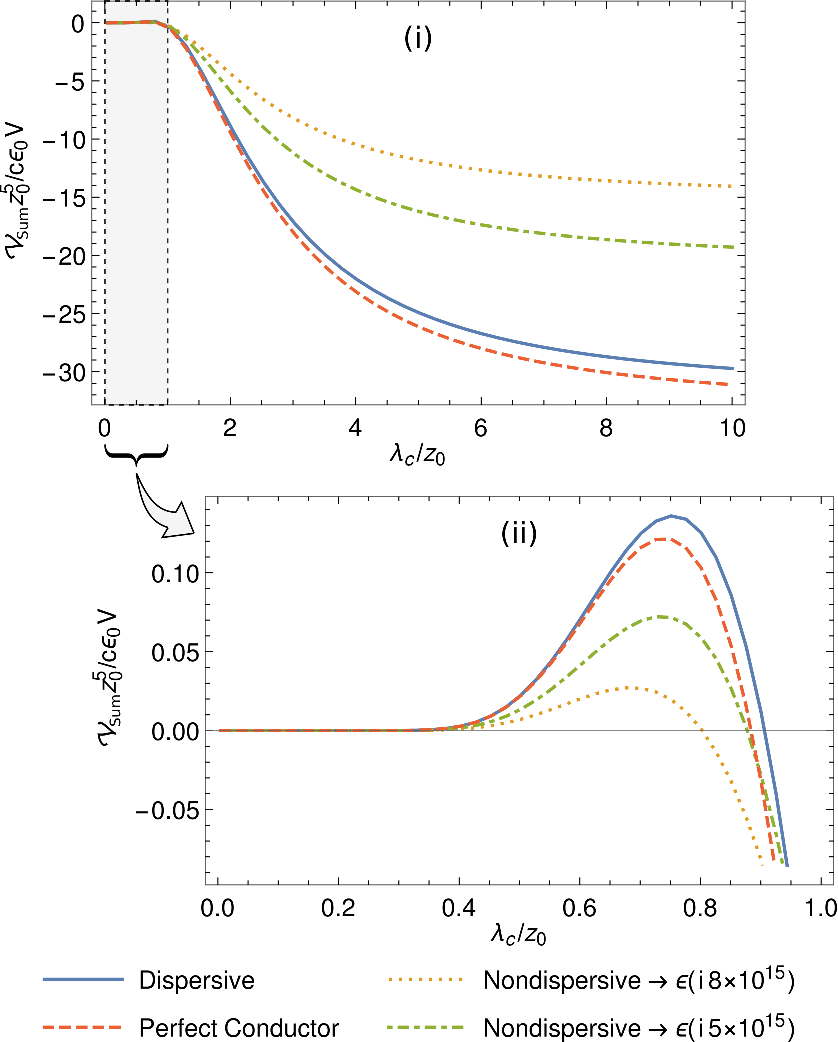, width=0.9 \linewidth}
\caption{
Behavior of ${\cal V}_{\text{Sum}}$ versus $\lambda_c/z_0$  for a gold spheroidal nanoparticle interacting with a gold sinusoidal surface computed using Eqs. \eqref{eq:vxx}-\eqref{eq:vzz}.
The particle is oriented such that its axis is parallel to the $x$-axis, and $\alpha_{mn}$ is described by Eqs. \eqref{eq:alpha-xx}-\eqref{eq:alpha-yz} with $\theta=\pi/2$ and $\phi=0$.
The highlighted interval in panel (i) is shown in panel (ii) in a zoomed in view.
In each panel it is considered $z_0 = 1 \, \mu\text{m}$ and $r = 2$, and it is shown the cases where the surface is considered dispersive (solid line), perfectly conductive (dashed line) and nondispersive [dotted line for $\epsilon(i 8 \times 10^{15})$ and dot-dashed for $\epsilon(i 5 \times 10^{15})$].
We remark that we have valley regime when ${\cal V}_{\text{Sum}} > 0$, and peak regime when ${\cal V}_{\text{Sum}} < 0$.
}
\label{fig:efeito-dispersao-CP}
\end{figure}

It is also interesting to perform the investigations above in the long distance case.
For this, in Fig. \ref{fig:efeito-dispersao-CP}, we consider Eqs. \eqref{eq:vxx}-\eqref{eq:vzz} for the particle oriented with its axis parallel to the $x$-axis, and we show the behavior of ${\cal V}_{\text{Sum}}$, in terms of $\lambda_c/z_0$, for the cases in which the corrugated surface is considered dispersive, nondispersive and perfectly conductive.
Similar to the vdW case, in this figure we have that the general behavior of ${\cal V}_{\text{Sum}}$ is similar for all cases, differing only numerically.
Moreover, it is known that for long distances from the surface the smallest frequencies give the main contribution to the interaction and, thus, $\epsilon(i\xi)$ can be replaced to its zero-frequency value in Eqs. \eqref{eq:vxx}-\eqref{eq:vzz}.
Because of this, the magnitudes of ${\cal V}_{\text{Sum}}$ for a dispersive surface are close to those for a perfect conductor (thus, are also greater than those for a nondispersive surface), as expected, since $\epsilon(0) \to \infty$ for a metallic surface.
Despite this, in the long distances case, we also have that the consideration of realistic material properties for the surface can amplify the occurrence of valley regime, since ${\cal V}_{\text{Sum}}$ changes its sign at a greater value of $\lambda_c/z_0$ for the dispersive case than the other cases.
Thus, although the consideration of realistic material properties for the surface weakens the interaction, it can increase the range of $\lambda_c/z_0$ values in which the valley regime occurs for both short and long distances cases.

\begin{figure}[h]
\centering
\epsfig{file=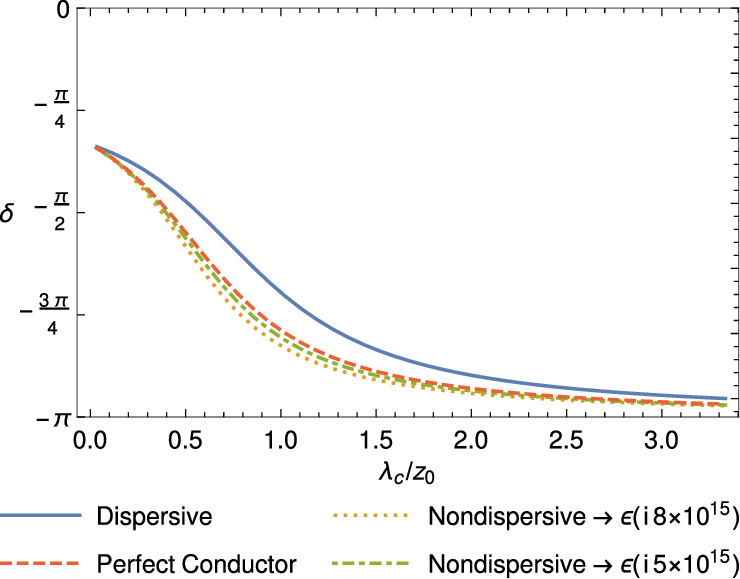, width=0.9 \linewidth}
\caption{
Behavior of $\delta$ versus $\lambda_c/z_0$  for a gold spheroidal nanoparticle interacting with a gold sinusoidal surface within the vdW regime [Eqs. \eqref{eq:vxx-vdw}-\eqref{eq:vzz-vdw}].
The particle is oriented such that $\alpha_{mn}$ is described by Eqs. \eqref{eq:alpha-xx}-\eqref{eq:alpha-yz} with $\theta=\pi/3$ and $\phi=0$ (its axis is in the $xz$-plane).
We consider $z_0 = 30 \text{ nm}$ and $r = 2$, and it is shown the cases where the surface is considered dispersive (solid line), perfectly conductive (dashed line) and nondispersive [dotted line for $\epsilon(i 8 \times 10^{15})$ and dot-dashed for $\epsilon(i 5 \times 10^{15})$].
}
\label{fig:efeito-dispersao-vdW-delta}
\end{figure}

The same previous discussions can also be made for the intermediate regimes, and we obtain the same conclusions.
To illustrate this, let us consider the particle oriented such that $\alpha_{mn}$ is described by Eqs. \eqref{eq:alpha-xx}-\eqref{eq:alpha-yz} with $\theta=\pi/3$ and $\phi=0$ (its axis is oriented in the $xz$-plane).
In this situation, $\alpha_{xz} \neq 0$, and we have intermediate regimes.
In Fig. \ref{fig:efeito-dispersao-vdW-delta} we show the behavior of the function $\delta$ [Eq. \ref{eq:funcao-delta}], versus $\lambda_c/z_0$, within the vdW regime for the cases in which the surface is considered dispersive, nondispersive and perfectly conductive.
Remembering that $\delta=0,\pi$ means valley and peak regimes, respectively, it is noted that when the surface is considered dispersive, the minimum points of $U^{(1)}$ tend to be further away from the corrugation peaks than in the other cases.
Thus, the consideration of realistic material properties for the surface also can amplify the occurrence of intermediate regimes.
Although we made this analysis within the vdW regime, we remark that the same conclusions are obtained for the long distance case.

\begin{figure}[h]
\centering
\subfigure[]{\label{fig:efeito-retardamento-05}\epsfig{file=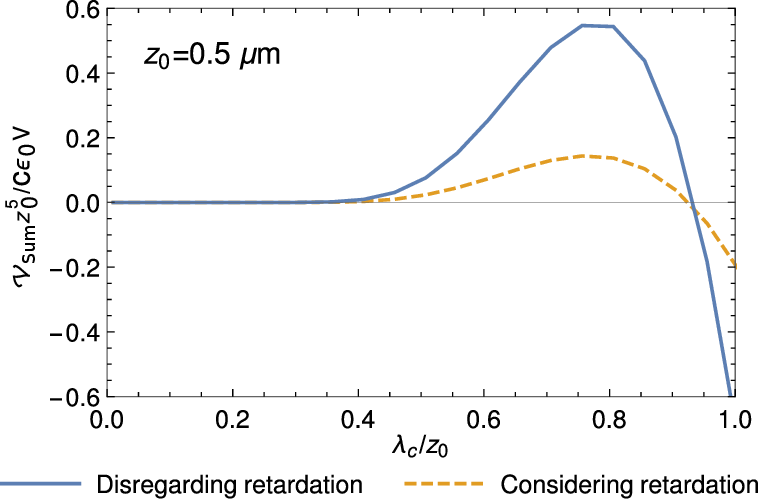, width=0.9 \linewidth}}
\subfigure[]{\label{fig:efeito-retardamento-10}\epsfig{file=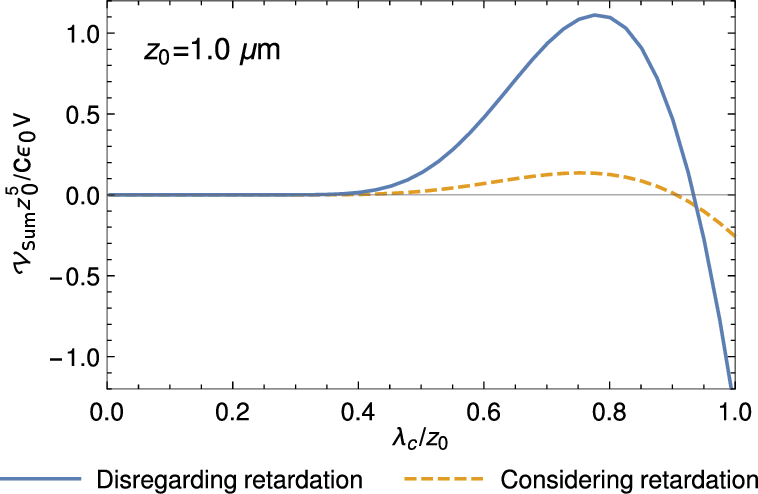, width=0.9 \linewidth}}
\caption{
Behavior of ${\cal V}_{\text{Sum}}$ versus $\lambda_c/z_0$  for a gold spheroidal nanoparticle interacting with a gold sinusoidal surface.
In both figures, the solid line shows this behavior disregarding retardation effects, i.e., using Eqs. \eqref{eq:vxx-vdw}-\eqref{eq:vzz-vdw}, while the dashed line shows it considering these effects by using Eqs. \eqref{eq:vxx}-\eqref{eq:vzz}.
In addition, we consider a particle with $r = 2$ and oriented such that its axis is parallel to the $x$-axis, so that $\alpha_{mn}$ is described by Eqs. \eqref{eq:alpha-xx}-\eqref{eq:alpha-yz} with $\theta=\pi/2$ and $\phi=0$.
In (a) we have $z_0 = 0.5 \mu\text{m}$, while in (b) $z_0 = 1.0 \mu\text{m}$.
We remark that we have valley regime when ${\cal V}_{\text{Sum}} > 0$, and peak regime when ${\cal V}_{\text{Sum}} < 0$.
}
\label{fig:efeito-retardamento}
\end{figure}

The investigation of the effect of retardation in the discussed interaction can be made by considering the particle at a long distance from the surface and comparing the results obtained from Eqs. \eqref{eq:vxx}-\eqref{eq:vxz} with those obtained from Eqs. \eqref{eq:vxx-vdw}-\eqref{eq:vxz-vdw}.
In this way, let us consider, again, the particle oriented such that its axis is parallel to the $x$-axis, and $\alpha_{mn}$ is described by Eqs. \eqref{eq:alpha-xx}-\eqref{eq:alpha-yz} with $\theta=\pi/2$ and $\phi=0$.
As previously discussed, in this situation we have valley regime when ${\cal V}_{\text{Sum}} > 0$, or peak regime when ${\cal V}_{\text{Sum}} < 0$.
In Fig. \ref{fig:efeito-retardamento}, we consider two values for $z_0$ outside the vdW regime and we show the behaviors of ${\cal V}_{\text{Sum}}$ obtained from Eqs. \eqref{eq:vxx}-\eqref{eq:vzz} and from Eqs. \eqref{eq:vxx-vdw}-\eqref{eq:vzz-vdw}.
Note that, as expected, the consideration of retardation weakens the interaction.
On the other hand, it has a small effect on the occurrence of valley regime, since ${\cal V}_{\text{Sum}}$ changes its sign by a slightly smaller value of $\lambda_c/z_0$ in both cases shown in Fig. \ref{fig:efeito-retardamento}.
This means that, when the goal is to estimate the value of $\lambda_c/z_0$ in which we have a transition between peak and valley regimes, the formulas within the vdW regime [i.e., Eqs. \eqref{eq:vxx-vdw}-\eqref{eq:vzz-vdw}] provide good approximate results.
Otherwise, when the goal is in obtaining precise values of the magnitudes of the energy, Eqs. \eqref{eq:vxx}-\eqref{eq:vzz} must be used.

\begin{figure}[h]
\centering
\epsfig{file=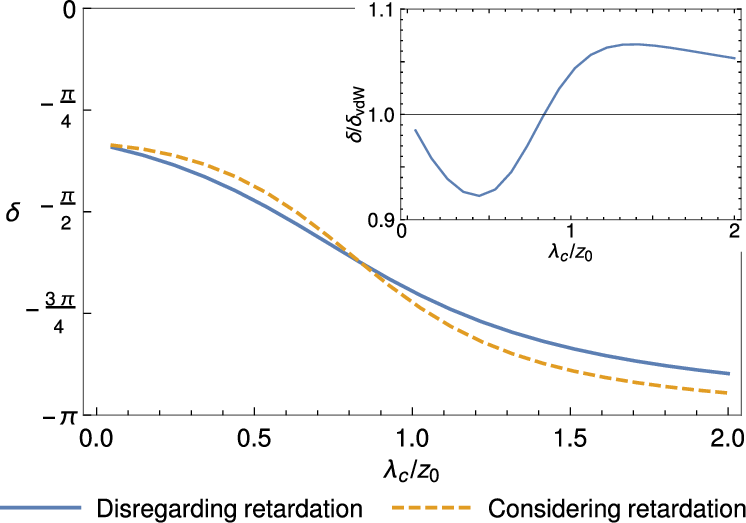, width=1 \linewidth}
\caption{
Behavior of $\delta$ versus $\lambda_c/z_0$ for a gold spheroidal nanoparticle interacting with a gold sinusoidal surface.
The solid line shows this behavior disregarding retardation effects, i.e., using Eqs. \eqref{eq:vxx-vdw}-\eqref{eq:vzz-vdw}, while the dashed one shows it considering these effects by using Eqs. \eqref{eq:vxx}-\eqref{eq:vzz}.
In addition, we consider the particle at $z_0 = 1.0 \mu\text{m}$, with $r = 2$, and oriented such that $\alpha_{mn}$ is described by Eqs. \eqref{eq:alpha-xx}-\eqref{eq:alpha-yz} with $\theta=\pi/3$ and $\phi=0$ (its axis is in the $xz$-axis).
The inset shows the ratio between the results calculated considering retardation effects and those disregarding them.
}
\label{fig:efeito-retardamento-delta}
\end{figure}

We also investigate the effects of retardation on the occurrence of the intermediate regimes.
This is shown in Fig. \ref{fig:efeito-retardamento-delta}, where we consider the particle at a distance from the surface outside the vdW regime and we show the behaviors of $\delta$ obtained from Eqs. \eqref{eq:vxx}-\eqref{eq:vxz} and from Eqs. \eqref{eq:vxx-vdw}-\eqref{eq:vxz-vdw}.
We also consider the particle oriented such that $\alpha_{mn}$ is described by Eqs. \eqref{eq:alpha-xx}-\eqref{eq:alpha-yz} with $\theta=\pi/3$ and $\phi=0$ (its axis is oriented in the $xz$-plane).
Remembering that $\delta=0,\pi$ means valley and peak regimes, respectively, it is noted in Fig. \ref{fig:efeito-retardamento-delta} that the consideration of retardation can amplify the occurrence of the intermediate regimes up to a certain value of $\lambda_c/z_0$, above which it begins to inhibit the effect.
This is reinforced in the inset of Fig. \ref{fig:efeito-retardamento-delta}, where we show the ratio between the results calculated considering retardation effects and those disregarding them.
%
%
%
%

%
\section{Final remarks}
\label{sec-final}

By means of the scattering approach, we obtained the Casimir-Polder interaction between a neutral anisotropic polarizable particle and a corrugated surface.
Focusing on the lateral force, which arises due to the presence of corrugation on the surface, we investigated how the occurrence of the peak, valley and intermediate regimes is affected by the consideration of realistic dielectric properties for the surface and of retardation on the interaction.
In this context, the discussions made here go beyond those made in Refs. \cite{Nogueira-PRA-2021,Queiroz-PRA-2021}, which considered idealized situations only in the vdW regime and also only obtained preliminary insights about the behavior of the mentioned effects.

Our main results are given by Eqs. \eqref{eq:u1}-\eqref{eq:funcao-amn}, whose application to a sinusoidal corrugation results in Eqs. \eqref{eq:u1-senoidal}-\eqref{eq:funcao-delta}, with the functions ${\cal V}_{mn}$ given by Eqs. \eqref{eq:vxx}-\eqref{eq:vxz}.
These equations, which take into account realistic properties for the surface and retardation effects in the interaction, not only generalize those found in Refs. \cite{Nogueira-PRA-2021, Queiroz-PRA-2021}, but also generalize the corresponding ones found in Ref. \cite{Messina-PRA-2009}, since we consider an anisotropic electrically polarizable particle.
Moreover, we showed that the consideration of realistic material properties for the surface, when compared to idealized materials (perfectly conductive or nondispersive), amplifies the occurrence of the valley and intermediate regimes.
Specifically, it increases the range of the values of $\lambda_c/z_0$ in which the valley regime occurs, or, when we have intermediate regimes, it shifts the minimum points of $U^{(1)}$ further away from the corrugation peaks (see Figs. \ref{fig:efeito-dispersao-vdW}-\ref{fig:efeito-dispersao-vdW-delta}).
In addition, we also showed that the effect of retardation on the occurrence of the valley regime is small (see Fig. \ref{fig:efeito-retardamento}), indicating that our vdW formulas [Eqs. \eqref{eq:vxx-vdw}-\eqref{eq:vxz-vdw}] provide good approximate results when the interest is in estimating the value of $\lambda_c/z_0$ in which we have a transition between peak and valley regimes.
On the occurrence of intermediate regimes, however, the consideration of retardation can amplify or inhibit it, which depends on the value of $\lambda_c/z_0$ (see Fig. \ref{fig:efeito-retardamento-delta}).
The results obtained here, which consider more realistic aspects for the particle-surface interaction, provide a preciser description of the interaction between an anisotropic particle and a corrugated surface, giving a better understanding of the nontrivial aspects of the lateral Casimir-Polder force.

\begin{acknowledgments}
The author expresses sincere gratitude to Danilo T. Alves for the careful reading of this paper and for the valuable orientation throughout the PhD.
The author also thanks Paulo A. Maia Neto for valuable discussions and comments.
This study was financed in part by the Coordena\c{c}\~{a}o de Aperfei\c{c}oamento de Pessoal de N\'{i}vel Superior - Brasil (CAPES) - Finance Code 001.
\end{acknowledgments}
%



\begin{thebibliography}{21}%
\makeatletter
\providecommand \@ifxundefined [1]{%
 \@ifx{#1\undefined}
}%
\providecommand \@ifnum [1]{%
 \ifnum #1\expandafter \@firstoftwo
 \else \expandafter \@secondoftwo
 \fi
}%
\providecommand \@ifx [1]{%
 \ifx #1\expandafter \@firstoftwo
 \else \expandafter \@secondoftwo
 \fi
}%
\providecommand \natexlab [1]{#1}%
\providecommand \enquote  [1]{``#1''}%
\providecommand \bibnamefont  [1]{#1}%
\providecommand \bibfnamefont [1]{#1}%
\providecommand \citenamefont [1]{#1}%
\providecommand \href@noop [0]{\@secondoftwo}%
\providecommand \href [0]{\begingroup \@sanitize@url \@href}%
\providecommand \@href[1]{\@@startlink{#1}\@@href}%
\providecommand \@@href[1]{\endgroup#1\@@endlink}%
\providecommand \@sanitize@url [0]{\catcode `\\12\catcode `\$12\catcode
  `\&12\catcode `\#12\catcode `\^12\catcode `\_12\catcode `\%12\relax}%
\providecommand \@@startlink[1]{}%
\providecommand \@@endlink[0]{}%
\providecommand \url  [0]{\begingroup\@sanitize@url \@url }%
\providecommand \@url [1]{\endgroup\@href {#1}{\urlprefix }}%
\providecommand \urlprefix  [0]{URL }%
\providecommand \Eprint [0]{\href }%
\providecommand \doibase [0]{https://doi.org/}%
\providecommand \selectlanguage [0]{\@gobble}%
\providecommand \bibinfo  [0]{\@secondoftwo}%
\providecommand \bibfield  [0]{\@secondoftwo}%
\providecommand \translation [1]{[#1]}%
\providecommand \BibitemOpen [0]{}%
\providecommand \bibitemStop [0]{}%
\providecommand \bibitemNoStop [0]{.\EOS\space}%
\providecommand \EOS [0]{\spacefactor3000\relax}%
\providecommand \BibitemShut  [1]{\csname bibitem#1\endcsname}%
\let\auto@bib@innerbib\@empty
\bibitem [{\citenamefont {Casimir}\ and\ \citenamefont
  {Polder}(1948)}]{Casimir-Polder-PhysRev-1948}%
  \BibitemOpen
  \bibfield  {author} {\bibinfo {author} {\bibfnamefont {H.~B.~G.}\
  \bibnamefont {Casimir}}\ and\ \bibinfo {author} {\bibfnamefont
  {D.}~\bibnamefont {Polder}},\ }\bibfield  {title} {\bibinfo {title} {{The
  Influence of Retardation on the {L}ondon-van der {W}aals Forces}},\ }\href
  {https://doi.org/10.1103/PhysRev.73.360} {\bibfield  {journal} {\bibinfo
  {journal} {Phys. Rev.}\ }\textbf {\bibinfo {volume} {73}},\ \bibinfo {pages}
  {360} (\bibinfo {year} {1948})}\BibitemShut {NoStop}%
\bibitem [{\citenamefont {Milonni}(1994)}]{Milonni-QuantumVacuum-1994}%
  \BibitemOpen
  \bibfield  {author} {\bibinfo {author} {\bibfnamefont {P.~W.}\ \bibnamefont
  {Milonni}},\ }\href@noop {} {\emph {\bibinfo {title} {{The Quantum Vacuum. An
  Introduction to Quantum Electrodynamics}}}}\ (\bibinfo  {publisher} {Academic
  Press},\ \bibinfo {address} {San Diego},\ \bibinfo {year} {1994})\BibitemShut
  {NoStop}%
\bibitem [{\citenamefont {Levin}\ \emph {et~al.}(2010)\citenamefont {Levin},
  \citenamefont {McCauley}, \citenamefont {Rodriguez}, \citenamefont {Reid},\
  and\ \citenamefont {Johnson}}]{Levin-PRL-2010}%
  \BibitemOpen
  \bibfield  {author} {\bibinfo {author} {\bibfnamefont {M.}~\bibnamefont
  {Levin}}, \bibinfo {author} {\bibfnamefont {A.~P.}\ \bibnamefont {McCauley}},
  \bibinfo {author} {\bibfnamefont {A.~W.}\ \bibnamefont {Rodriguez}}, \bibinfo
  {author} {\bibfnamefont {M.~T.~H.}\ \bibnamefont {Reid}},\ and\ \bibinfo
  {author} {\bibfnamefont {S.~G.}\ \bibnamefont {Johnson}},\ }\bibfield
  {title} {\bibinfo {title} {Casimir repulsion between metallic objects in
  vacuum},\ }\href {https://doi.org/10.1103/PhysRevLett.105.090403} {\bibfield
  {journal} {\bibinfo  {journal} {Phys. Rev. Lett.}\ }\textbf {\bibinfo
  {volume} {105}},\ \bibinfo {pages} {090403} (\bibinfo {year}
  {2010})}\BibitemShut {NoStop}%
\bibitem [{\citenamefont {Eberlein}\ and\ \citenamefont
  {Zietal}(2011)}]{Eberlein-PRA-2011}%
  \BibitemOpen
  \bibfield  {author} {\bibinfo {author} {\bibfnamefont {C.}~\bibnamefont
  {Eberlein}}\ and\ \bibinfo {author} {\bibfnamefont {R.}~\bibnamefont
  {Zietal}},\ }\bibfield  {title} {\bibinfo {title} {{Casimir-{P}older
  interaction between a polarizable particle and a plate with a hole}},\ }\href
  {https://doi.org/10.1103/PhysRevA.83.052514} {\bibfield  {journal} {\bibinfo
  {journal} {Phys. Rev. A}\ }\textbf {\bibinfo {volume} {83}},\ \bibinfo
  {pages} {052514} (\bibinfo {year} {2011})}\BibitemShut {NoStop}%
\bibitem [{\citenamefont {Buhmann}\ \emph {et~al.}(2016)\citenamefont
  {Buhmann}, \citenamefont {Marachevsky},\ and\ \citenamefont
  {Scheel}}]{Buhmann-IJMPA-2016}%
  \BibitemOpen
  \bibfield  {author} {\bibinfo {author} {\bibfnamefont {S.~Y.}\ \bibnamefont
  {Buhmann}}, \bibinfo {author} {\bibfnamefont {V.~N.}\ \bibnamefont
  {Marachevsky}},\ and\ \bibinfo {author} {\bibfnamefont {S.}~\bibnamefont
  {Scheel}},\ }\bibfield  {title} {\bibinfo {title} {{Impact of anisotropy on
  the interaction of an atom with a one-dimensional nano-grating}},\ }\href
  {https://doi.org/10.1142/S0217751X16410293} {\bibfield  {journal} {\bibinfo
  {journal} {Int. J. Mod. Phys. A}\ }\textbf {\bibinfo {volume} {31}},\
  \bibinfo {pages} {1641029} (\bibinfo {year} {2016})}\BibitemShut {NoStop}%
\bibitem [{\citenamefont {Abrantes}\ \emph {et~al.}(2018)\citenamefont
  {Abrantes}, \citenamefont {Fran{\c{c}}a}, \citenamefont {da~Rosa},
  \citenamefont {Farina},\ and\ \citenamefont {{de Melo e
  Souza}}}]{Abrantes-PRA-2018}%
  \BibitemOpen
  \bibfield  {author} {\bibinfo {author} {\bibfnamefont {P.~P.}\ \bibnamefont
  {Abrantes}}, \bibinfo {author} {\bibfnamefont {Y.}~\bibnamefont
  {Fran{\c{c}}a}}, \bibinfo {author} {\bibfnamefont {F.~S.~S.}\ \bibnamefont
  {da~Rosa}}, \bibinfo {author} {\bibfnamefont {C.}~\bibnamefont {Farina}},\
  and\ \bibinfo {author} {\bibfnamefont {R.}~\bibnamefont {{de Melo e
  Souza}}},\ }\bibfield  {title} {\bibinfo {title} {{Repulsive van der {W}aals
  interaction between a quantum particle and a conducting toroid}},\ }\href
  {https://doi.org/10.1103/PhysRevA.98.012511} {\bibfield  {journal} {\bibinfo
  {journal} {Phys. Rev. A}\ }\textbf {\bibinfo {volume} {98}},\ \bibinfo
  {pages} {012511} (\bibinfo {year} {2018})}\BibitemShut {NoStop}%
\bibitem [{\citenamefont {Venkataram}\ \emph {et~al.}(2020)\citenamefont
  {Venkataram}, \citenamefont {Molesky}, \citenamefont {Chao},\ and\
  \citenamefont {Rodriguez}}]{Venkataram-PRA-2020}%
  \BibitemOpen
  \bibfield  {author} {\bibinfo {author} {\bibfnamefont {P.~S.}\ \bibnamefont
  {Venkataram}}, \bibinfo {author} {\bibfnamefont {S.}~\bibnamefont {Molesky}},
  \bibinfo {author} {\bibfnamefont {P.}~\bibnamefont {Chao}},\ and\ \bibinfo
  {author} {\bibfnamefont {A.~W.}\ \bibnamefont {Rodriguez}},\ }\bibfield
  {title} {\bibinfo {title} {{Fundamental limits to attractive and repulsive
  {C}asimir-{P}older forces}},\ }\href
  {https://doi.org/10.1103/PhysRevA.101.052115} {\bibfield  {journal} {\bibinfo
   {journal} {Phys. Rev. A}\ }\textbf {\bibinfo {volume} {101}},\ \bibinfo
  {pages} {052115} (\bibinfo {year} {2020})}\BibitemShut {NoStop}%
\bibitem [{\citenamefont {Marchetta}\ \emph {et~al.}(2021)\citenamefont
  {Marchetta}, \citenamefont {Parashar},\ and\ \citenamefont
  {Shajesh}}]{Marchetta-PRA-2021}%
  \BibitemOpen
  \bibfield  {author} {\bibinfo {author} {\bibfnamefont {J.~J.}\ \bibnamefont
  {Marchetta}}, \bibinfo {author} {\bibfnamefont {P.}~\bibnamefont
  {Parashar}},\ and\ \bibinfo {author} {\bibfnamefont {K.~V.}\ \bibnamefont
  {Shajesh}},\ }\bibfield  {title} {\bibinfo {title} {{Geometrical dependence
  in {C}asimir-{P}older repulsion}},\ }\href
  {https://doi.org/10.1103/PhysRevA.104.032209} {\bibfield  {journal} {\bibinfo
   {journal} {Phys. Rev. A}\ }\textbf {\bibinfo {volume} {104}},\ \bibinfo
  {pages} {032209} (\bibinfo {year} {2021})}\BibitemShut {NoStop}%
\bibitem [{\citenamefont {Bimonte}\ \emph {et~al.}(2015)\citenamefont
  {Bimonte}, \citenamefont {Emig},\ and\ \citenamefont
  {Kardar}}]{Bimonte-PRD-2015}%
  \BibitemOpen
  \bibfield  {author} {\bibinfo {author} {\bibfnamefont {G.}~\bibnamefont
  {Bimonte}}, \bibinfo {author} {\bibfnamefont {T.}~\bibnamefont {Emig}},\ and\
  \bibinfo {author} {\bibfnamefont {M.}~\bibnamefont {Kardar}},\ }\bibfield
  {title} {\bibinfo {title} {{Casimir-{P}older force between anisotropic
  nanoparticles and gently curved surfaces}},\ }\href
  {https://doi.org/10.1103/PhysRevD.92.025028} {\bibfield  {journal} {\bibinfo
  {journal} {Phys. Rev. D}\ }\textbf {\bibinfo {volume} {92}},\ \bibinfo
  {pages} {025028} (\bibinfo {year} {2015})}\BibitemShut {NoStop}%
\bibitem [{\citenamefont {Gangaraj}\ \emph {et~al.}(2018)\citenamefont
  {Gangaraj}, \citenamefont {Silveirinha}, \citenamefont {Hanson},
  \citenamefont {Antezza},\ and\ \citenamefont
  {Monticone}}]{Gangaraj-PRB-2018}%
  \BibitemOpen
  \bibfield  {author} {\bibinfo {author} {\bibfnamefont {S.~A.~H.}\
  \bibnamefont {Gangaraj}}, \bibinfo {author} {\bibfnamefont {M.~G.}\
  \bibnamefont {Silveirinha}}, \bibinfo {author} {\bibfnamefont {G.~W.}\
  \bibnamefont {Hanson}}, \bibinfo {author} {\bibfnamefont {M.}~\bibnamefont
  {Antezza}},\ and\ \bibinfo {author} {\bibfnamefont {F.}~\bibnamefont
  {Monticone}},\ }\bibfield  {title} {\bibinfo {title} {{Optical torque on a
  two-level system near a strongly nonreciprocal medium}},\ }\href
  {https://doi.org/10.1103/PhysRevB.98.125146} {\bibfield  {journal} {\bibinfo
  {journal} {Phys. Rev. B}\ }\textbf {\bibinfo {volume} {98}},\ \bibinfo
  {pages} {125146} (\bibinfo {year} {2018})}\BibitemShut {NoStop}%
\bibitem [{\citenamefont {Antezza}\ \emph {et~al.}(2020)\citenamefont
  {Antezza}, \citenamefont {Fialkovsky},\ and\ \citenamefont
  {Khusnutdinov}}]{Antezza-PRB-2020}%
  \BibitemOpen
  \bibfield  {author} {\bibinfo {author} {\bibfnamefont {M.}~\bibnamefont
  {Antezza}}, \bibinfo {author} {\bibfnamefont {I.}~\bibnamefont
  {Fialkovsky}},\ and\ \bibinfo {author} {\bibfnamefont {N.}~\bibnamefont
  {Khusnutdinov}},\ }\bibfield  {title} {\bibinfo {title} {{Casimir-{P}older
  force and torque for anisotropic molecules close to conducting planes and
  their effect on ${\text{CO}}_{2}$}},\ }\href
  {https://doi.org/10.1103/PhysRevB.102.195422} {\bibfield  {journal} {\bibinfo
   {journal} {Phys. Rev. B}\ }\textbf {\bibinfo {volume} {102}},\ \bibinfo
  {pages} {195422} (\bibinfo {year} {2020})}\BibitemShut {NoStop}%
\bibitem [{\citenamefont {Nogueira}\ \emph {et~al.}(2021)\citenamefont
  {Nogueira}, \citenamefont {Queiroz},\ and\ \citenamefont
  {Alves}}]{Nogueira-PRA-2021}%
  \BibitemOpen
  \bibfield  {author} {\bibinfo {author} {\bibfnamefont {E.~C.~M.}\
  \bibnamefont {Nogueira}}, \bibinfo {author} {\bibfnamefont {L.}~\bibnamefont
  {Queiroz}},\ and\ \bibinfo {author} {\bibfnamefont {D.~T.}\ \bibnamefont
  {Alves}},\ }\bibfield  {title} {\bibinfo {title} {Peak, valley, and
  intermediate regimes in the lateral van der {W}aals force},\ }\href
  {https://doi.org/10.1103/PhysRevA.104.012816} {\bibfield  {journal} {\bibinfo
   {journal} {Phys. Rev. A}\ }\textbf {\bibinfo {volume} {104}},\ \bibinfo
  {pages} {012816} (\bibinfo {year} {2021})}\BibitemShut {NoStop}%
\bibitem [{\citenamefont {Queiroz}\ \emph {et~al.}(2021)\citenamefont
  {Queiroz}, \citenamefont {Nogueira},\ and\ \citenamefont
  {Alves}}]{Queiroz-PRA-2021}%
  \BibitemOpen
  \bibfield  {author} {\bibinfo {author} {\bibfnamefont {L.}~\bibnamefont
  {Queiroz}}, \bibinfo {author} {\bibfnamefont {E.~C.~M.}\ \bibnamefont
  {Nogueira}},\ and\ \bibinfo {author} {\bibfnamefont {D.~T.}\ \bibnamefont
  {Alves}},\ }\bibfield  {title} {\bibinfo {title} {{Regimes of the lateral van
  der Waals force in the presence of dielectrics}},\ }\href
  {https://doi.org/10.1103/PhysRevA.104.062802} {\bibfield  {journal} {\bibinfo
   {journal} {Phys. Rev. A}\ }\textbf {\bibinfo {volume} {104}},\ \bibinfo
  {pages} {062802} (\bibinfo {year} {2021})}\BibitemShut {NoStop}%
\bibitem [{\citenamefont {Nogueira}\ \emph {et~al.}(2022)\citenamefont
  {Nogueira}, \citenamefont {Queiroz},\ and\ \citenamefont
  {Alves}}]{Nogueira-PRA-2022}%
  \BibitemOpen
  \bibfield  {author} {\bibinfo {author} {\bibfnamefont {E.~C.~M.}\
  \bibnamefont {Nogueira}}, \bibinfo {author} {\bibfnamefont {L.}~\bibnamefont
  {Queiroz}},\ and\ \bibinfo {author} {\bibfnamefont {D.~T.}\ \bibnamefont
  {Alves}},\ }\bibfield  {title} {\bibinfo {title} {{Sign inversion in the
  lateral van der Waals force}},\ }\href
  {https://doi.org/10.1103/PhysRevA.105.062816} {\bibfield  {journal} {\bibinfo
   {journal} {Phys. Rev. A}\ }\textbf {\bibinfo {volume} {105}},\ \bibinfo
  {pages} {062816} (\bibinfo {year} {2022})}\BibitemShut {NoStop}%
\bibitem [{\citenamefont {Queiroz}\ \emph {et~al.}(2023)\citenamefont
  {Queiroz}, \citenamefont {Nogueira},\ and\ \citenamefont
  {Alves}}]{Queiroz-JPA-2023}%
  \BibitemOpen
  \bibfield  {author} {\bibinfo {author} {\bibfnamefont {L.}~\bibnamefont
  {Queiroz}}, \bibinfo {author} {\bibfnamefont {E.~C.~M.}\ \bibnamefont
  {Nogueira}},\ and\ \bibinfo {author} {\bibfnamefont {D.~T.}\ \bibnamefont
  {Alves}},\ }\bibfield  {title} {\bibinfo {title} {Sign inversion in the
  lateral van der {W}aals force between an anisotropic particle and a plane
  with a hemispherical protuberance: an exact calculation},\ }\href
  {https://doi.org/10.1088/1751-8121/acb4c7} {\bibfield  {journal} {\bibinfo
  {journal} {J. Phys. A: Math. Theor.}\ }\textbf
  {\bibinfo {volume} {56}},\ \bibinfo {pages} {115301} (\bibinfo {year}
  {2023})}\BibitemShut {NoStop}%
\bibitem [{\citenamefont {Alves}\ \emph {et~al.}(2023)\citenamefont {Alves},
  \citenamefont {Queiroz}, \citenamefont {Nogueira},\ and\ \citenamefont
  {Peres}}]{Alves-PRA-2023}%
  \BibitemOpen
  \bibfield  {author} {\bibinfo {author} {\bibfnamefont {D.~T.}\ \bibnamefont
  {Alves}}, \bibinfo {author} {\bibfnamefont {L.}~\bibnamefont {Queiroz}},
  \bibinfo {author} {\bibfnamefont {E.~C.~M.}\ \bibnamefont {Nogueira}},\ and\
  \bibinfo {author} {\bibfnamefont {N.~M.~R.}\ \bibnamefont {Peres}},\
  }\bibfield  {title} {\bibinfo {title} {{Curvature-induced repulsive effect on
  the lateral Casimir-Polder–van der Waals force}},\ }\href
  {https://doi.org/10.1103/PhysRevA.107.062821} {\bibfield  {journal} {\bibinfo
   {journal} {Phys. Rev. A}\ }\textbf {\bibinfo {volume} {107}},\ \bibinfo
  {pages} {062821} (\bibinfo {year} {2023})}\BibitemShut {NoStop}%
\bibitem [{\citenamefont {Messina}\ \emph {et~al.}(2009)\citenamefont
  {Messina}, \citenamefont {Dalvit}, \citenamefont {Neto}, \citenamefont
  {Lambrecht},\ and\ \citenamefont {Reynaud}}]{Messina-PRA-2009}%
  \BibitemOpen
  \bibfield  {author} {\bibinfo {author} {\bibfnamefont {R.}~\bibnamefont
  {Messina}}, \bibinfo {author} {\bibfnamefont {D.~A.~R.}\ \bibnamefont
  {Dalvit}}, \bibinfo {author} {\bibfnamefont {P.~A.~M.}\ \bibnamefont {Neto}},
  \bibinfo {author} {\bibfnamefont {A.}~\bibnamefont {Lambrecht}},\ and\
  \bibinfo {author} {\bibfnamefont {S.}~\bibnamefont {Reynaud}},\ }\bibfield
  {title} {\bibinfo {title} {{Dispersive interactions between atoms and
  nonplanar surfaces}},\ }\href {https://doi.org/10.1103/PhysRevA.80.022119}
  {\bibfield  {journal} {\bibinfo  {journal} {Phys. Rev. A}\ }\textbf
  {\bibinfo {volume} {80}},\ \bibinfo {pages} {022119} (\bibinfo {year}
  {2009})}\BibitemShut {NoStop}%
\bibitem [{\citenamefont {Neto}\ \emph
  {et~al.}(2005{\natexlab{a}})\citenamefont {Neto}, \citenamefont {Lambrecht},\
  and\ \citenamefont {Reynaud}}]{Neto-EPL-2005}%
  \BibitemOpen
  \bibfield  {author} {\bibinfo {author} {\bibfnamefont {P.~A.~M.}\
  \bibnamefont {Neto}}, \bibinfo {author} {\bibfnamefont {A.}~\bibnamefont
  {Lambrecht}},\ and\ \bibinfo {author} {\bibfnamefont {S.}~\bibnamefont
  {Reynaud}},\ }\bibfield  {title} {\bibinfo {title} {{Roughness correction to
  the Casimir force: Beyond the Proximity Force Approximation}},\ }\href
  {https://doi.org/10.1209/epl/i2004-10433-9} {\bibfield  {journal} {\bibinfo
  {journal} {Europhys. Lett. (EPL)}\ }\textbf {\bibinfo {volume} {69}},\
  \bibinfo {pages} {924} (\bibinfo {year} {2005}{\natexlab{a}})}\BibitemShut
  {NoStop}%
\bibitem [{\citenamefont {Neto}\ \emph
  {et~al.}(2005{\natexlab{b}})\citenamefont {Neto}, \citenamefont {Lambrecht},\
  and\ \citenamefont {Reynaud}}]{Neto-PRA-2005}%
  \BibitemOpen
  \bibfield  {author} {\bibinfo {author} {\bibfnamefont {P.~A.~M.}\
  \bibnamefont {Neto}}, \bibinfo {author} {\bibfnamefont {A.}~\bibnamefont
  {Lambrecht}},\ and\ \bibinfo {author} {\bibfnamefont {S.}~\bibnamefont
  {Reynaud}},\ }\bibfield  {title} {\bibinfo {title} {{Casimir effect with
  rough metallic mirrors}},\ }\href
  {https://doi.org/10.1103/PhysRevA.72.012115} {\bibfield  {journal} {\bibinfo
  {journal} {Phys. Rev. A}\ }\textbf {\bibinfo {volume} {72}},\ \bibinfo
  {pages} {012115} (\bibinfo {year} {2005}{\natexlab{b}})}\BibitemShut
  {NoStop}%
\bibitem [{\citenamefont {Buhmann}(2012)}]{Buhmann-DispersionForces-II}%
  \BibitemOpen
  \bibfield  {author} {\bibinfo {author} {\bibfnamefont {S.~Y.}\ \bibnamefont
  {Buhmann}},\ }\href {https://doi.org/10.1007/978-3-642-32466-6} {\emph
  {\bibinfo {title} {{Dispersion Forces II}}}},\ \bibinfo {series} {Springer
  Tracts in Modern Physics}, Vol.\ \bibinfo {volume} {248}\ (\bibinfo
  {publisher} {Springer Berlin Heidelberg},\ \bibinfo {address} {Berlin,
  Heidelberg},\ \bibinfo {year} {2012})\BibitemShut {NoStop}%
\bibitem [{\citenamefont {Dalvit}\ \emph {et~al.}(2008)\citenamefont {Dalvit},
  \citenamefont {Neto}, \citenamefont {Lambrecht},\ and\ \citenamefont
  {Reynaud}}]{Dalvit-JPA-2008}%
  \BibitemOpen
  \bibfield  {author} {\bibinfo {author} {\bibfnamefont {D.~A.~R.}\
  \bibnamefont {Dalvit}}, \bibinfo {author} {\bibfnamefont {P.~A.~M.}\
  \bibnamefont {Neto}}, \bibinfo {author} {\bibfnamefont {A.}~\bibnamefont
  {Lambrecht}},\ and\ \bibinfo {author} {\bibfnamefont {S.}~\bibnamefont
  {Reynaud}},\ }\bibfield  {title} {\bibinfo {title} {{Lateral
  {C}asimir-{P}older force with corrugated surfaces}},\ }\href
  {https://doi.org/10.1088/1751-8113/41/16/164028} {\bibfield  {journal}
  {\bibinfo  {journal} {J. Phys. A: Math. Theor.}\
  }\textbf {\bibinfo {volume} {41}},\ \bibinfo {pages} {164028} (\bibinfo
  {year} {2008})}\BibitemShut {NoStop}%
\end{thebibliography}

%

\end{document}